\documentclass[%
 reprint,
superscriptaddress,
 amsmath,amssymb,
 aps,
]{revtex4-2}

\usepackage{graphicx}
\usepackage{dcolumn}
\usepackage{bm}
\usepackage[version=3]{mhchem} 
\usepackage{xcolor} 

\begin{document}

\preprint{APS/123-QED}

\title{
Probing the low-energy electron-scattering dynamics in liquids \\ with high-harmonic spectroscopy }

\author{Angana Mondal}
\thanks{These five authors contributed equally}
\affiliation{Laboratorium für Physikalische Chemie, ETH Zürich, Zurich, Switzerland}

\author{Ofer Neufeld}
\thanks{These five authors contributed equally}
\affiliation{Max Planck Institute for the Structure and Dynamics of Matter, Luruper Chaussee 149, 22761 Hamburg, Germany}
\affiliation{Center for Free-Electron Laser Science CFEL, Deutsches Elektronen-Synchrotron DESY, Notkestra\ss e 85, 22607 Hamburg, Germany}

\author{Zhong Yin}
\thanks{These five authors contributed equally}
\affiliation{Laboratorium für Physikalische Chemie, ETH Zürich, Zurich, Switzerland}

\author{Zahra Nourbakhsh}
\thanks{These five authors contributed equally}
\affiliation{Max Planck Institute for the Structure and Dynamics of Matter, Luruper Chaussee 149, 22761 Hamburg, Germany}
\affiliation{Center for Free-Electron Laser Science CFEL, Deutsches Elektronen-Synchrotron DESY, Notkestra\ss e 85, 22607 Hamburg, Germany}

\author{V\'it Svoboda}
\thanks{These five authors contributed equally}
\affiliation{Laboratorium für Physikalische Chemie, ETH Zürich, Zurich, Switzerland}

\author{Angel Rubio}
\affiliation{Center for Free-Electron Laser Science CFEL, Deutsches Elektronen-Synchrotron DESY, Notkestra\ss e 85, 22607 Hamburg, Germany}
\affiliation{Physics Department, University of Hamburg, Luruper Chaussee 149, 22761 Hamburg, Germany}
\affiliation{Max Planck Institute for the Structure and Dynamics of Matter, Luruper Chaussee 149, 22761 Hamburg, Germany}
\affiliation{The Hamburg Centre for Ultrafast Imaging, Luruper Chaussee 149, 22761 Hamburg, Germany}
\affiliation{Center for Computational Quantum Physics (CCQ), The Flatiron Institute, 162 Fifth Avenue, New York NY 10010, USA}

\author{Nicolas Tancogne-Dejean}
\affiliation{Max Planck Institute for the Structure and Dynamics of Matter, Luruper Chaussee 149, 22761 Hamburg, Germany}
\affiliation{Center for Free-Electron Laser Science CFEL, Deutsches Elektronen-Synchrotron DESY, Notkestra\ss e 85, 22607 Hamburg, Germany}

\author{Hans Jakob W\"orner}
\affiliation{Laboratorium für Physikalische Chemie, ETH Zürich, Zurich, Switzerland}

\begin{abstract}
High-harmonic spectroscopy (HHS) is a nonlinear all-optical technique with inherent attosecond temporal resolution, which has been applied successfully to a broad variety of systems in the gas phase and solid state. Here, we extend HHS to the liquid phase, and uncover the mechanism of high-harmonic generation (HHG) for this phase of matter. Studying HHG over a broad range of wavelengths and intensities, we show that the cut-off ($E_c$) is independent of the wavelength beyond a threshold intensity, and find that $E_c$ is a characteristic property of the studied liquid. We explain these observations within an intuitive semi-classical model based on electron trajectories that are limited by scattering to a characteristic length, which is connected to the electron mean-free path. This is further confirmed by measurements performed with elliptically polarized laser fields. Our model is validated against rigorous multi-electron time-dependent density-functional-theory calculations in, both, supercells of liquid water with periodic boundary conditions, and large clusters of a variety of liquids. These simulations confirm our interpretation and thereby clarify the mechanism of HHG in liquids. Our results demonstrate a new, all-optical indirect access to effective mean-free paths of slow electrons ($\leq$10~eV) in liquids, in a regime that is inaccessible to accurate calculations, but is critical for the understanding of radiation damage to living tissue. 
Our work also establishes the possibility of resolving sub-femtosecond electron dynamics in liquids, which offers a novel, all-optical approach to attosecond spectroscopy of chemical processes in their native liquid environment. 

\end{abstract}

\maketitle

\section{Introduction}

High-harmonic generation (HHG) is an extremely nonlinear process that occurs when a strong laser field interacts with gaseous, solid, or liquid targets. It results in an up-conversion of photon energies up to the tender X-ray regime \cite{Popmintchev2012}, and has been well established as a highly versatile table-top source of attosecond pulses \cite{Cousin2014, Teichmann2016, Pertot2017, Attar2017, Gaumnitz2017, Schmidt2018, Smith2020}. HHG has also led to a new branch of attosecond spectroscopy that relies on extracting dynamical information directly from measured spectra, known as high-harmonic spectroscopy (HHS). This technique has already enabled imaging of molecular orbitals \cite{itatani04a,Haessler2009,peng2019}, the reconstruction of charge migration \cite{Kraus2015} and time-dependent chirality \cite{baykusheva2019}, as well as tunneling-ionization dynamics \cite{Shafir2012}. While the majority of these applications are based on gas-phase HHG, solid-state HHG has recently attracted considerable attention because of its potentially higher efficiency and access to ultrafast dynamics and light-driven phase transitions in condensed matter \cite{ghimire11a,vampa2015,luu2015,ghimire2019}. A prerequisite for an accurate interpretation of the underlying dynamics from the measured high-harmonic spectra lies in the formulation of a broadly applicable theoretical model.

In the gas phase, this understanding and modelling is often based on the semi-classical three-step model (TSM) \cite{Corkum1993}, or its quantum-mechanical extension \cite{Lewenstein1994}, which describes HHG as a set of electron trajectories initiated by a tunneling process. This approach is usually in good agreement with full \textit{ab-initio} calculations, and allows interpretation of dynamical information in HHS \cite{Worner2010,Uzan2020}. A hallmark of the model is that it correctly predicts the HHG cut-off and its quadratic dependence on, both, electric-field amplitude and wavelength, connected to the most energetic returning electron trajectory \cite{Huillier1993, gordon2005}. For HHG in crystalline solids a similar trajectory picture, based on the material's band structure, can in principle be applied in momentum space (after applying the Bloch theorem) \cite{Ghimire2011, schubert2014, Osika2017, luu16a, Li2019, vampa2015, Wu2016, Yue2021}. The predicted cut-off energy was shown to scale linearly with both field amplitude and wavelength \cite{Ghimire2011, Colosimo2008, liu2018wavelength, luu2015, Liu2017, Navarrete2019, vampa2015, Wu2016}. However, there still remains some debate about the scaling based on the active HHG mechanisms in different solid systems \cite{ghimire2019, Tancogne2017, nor2021, li2021huygens}. Moreover, the direct comparison of crystalline and amorphous solids of the same composition (quartz vs. fused silica) has shown that the cut-off energy is much lower in the latter under the same driving fields \cite{luu18b}. Whereas this observation remains to be fully explained, it points to the importance of long-range order in condensed-phase HHG, which was also explored for one-dimensional models \cite{Yu2019,Zeng2020}.
\begin{figure*} [t!]
\includegraphics[width=\textwidth]{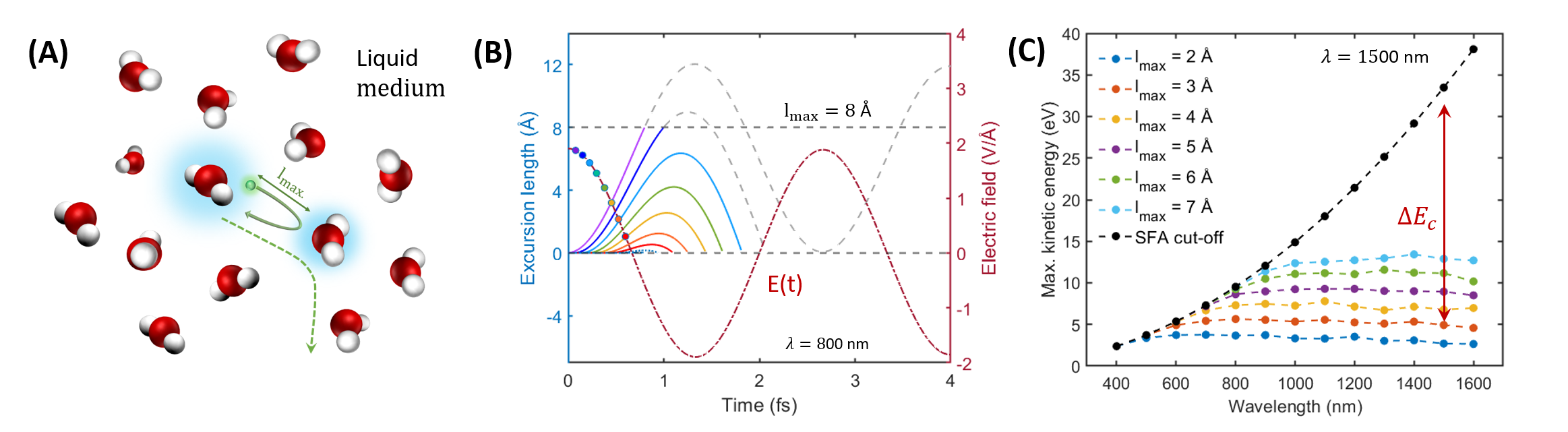}
\caption{{\bf Effect of electron scattering on HHG spectra in liquids}
\textbf{(A)} 
Schematic illustration of the extended trajectory-based semiclassical model. An electron (green) is ionized by the laser field, accelerated and either recombines directly with its parent ion (solid green arrows), or scatters off another molecule (dashed green arrows). 
\textbf{(B)} 
The returning electron trajectories from the standard TSM within an 800-nm driving laser field of 1.9 V$/$\AA. The dots on the electric field represent the ionization times of the electrons (the colour of dots corresponding to the colour of the respective trajectories). The grey dashed line denotes the limited excursion length (l$_{\rm max}$) imposed by scattering. 
\textbf{(C)} 
Wavelength-scaling of $E_c$ in the absence (standard TSM: black dashed line) and presence (l$_{\rm max}$-limited: colored lines) of scattering for a laser intensity of 5$\times$10$^{13}$ W/cm$^2$. The observable manifestation of scattering is a decrease in the cut-off by $\Delta E_c$.}
\label{fig:one}
\end{figure*}
In contrast to gases and crystals, HHG in the liquid phase is far from being well understood. This is because it is challenging to: (i) experimentally measure HHG spectra from bulk liquids, (ii) numerically simulate strong-field physics in liquids, and (iii) there is still no intuitive model that describes non-perturbative light-driven dynamics in liquids. Liquids, therefore, present a unique case where neither gas (single isolated particle approach) nor solid-state (Bloch theorem and periodic boundary conditions) approaches are strictly applicable. This gap in knowledge limits potential applications to ultrafast spectroscopy that are especially appealing in liquid targets. Only very recently, HHG in bulk liquids have been demonstrated beyond the visible domain utilizing the flat-jet approach \cite{Luu2018,yin2020,svoboda2021polarization}. However, fundamental questions about the dominant HHG microscopic mechanisms, the cut-off scaling with wavelength or the macroscopic effects still remain unanswered. As the major bio-chemical processes take place in a liquid environment, detailed experimental results and the development of theoretical tools capable of describing the HHG process are crucial for understanding the electron dynamics in liquids. We note that our work addresses HHG at typical intensities of $\sim 10^{13}$~W/cm$^2$, which is the basis of HHS, in contrast to HHG in the coherent-wake-emission regime taking place at intensities beyond 10$^{17}$W/cm$^2$. The latter has been demonstrated on the surface of liquids \cite{heissler14a}. As a consequence of the broken inversion symmetry, both even and odd harmonics were observed in those experiments. Our present experiments, in contrast, probe the bulk of the liquid phase, such that no even harmonics are observed.

Here, we experimentally measure high-harmonic spectra from liquid water and alcohols over a broad range of laser wavelengths. We observe that the HHG cut-off energy ($E_c$), i.e. the energy marking the end of plateau region, as defined in the Lewenstein formalism \cite{Lewenstein1994}, is wavelength independent in strong contrast with the semi-classical TSM for gases \cite{gordon2005}, as well as some models for solid-state HHG \cite{liu2018wavelength, Ghimire2011, schubert2014, Osika2017, Li2019, vampa2015, Wu2016, Yue2021}. This implies that potentially new mechanisms are relevant in liquid HHG, and that the structural arrangement of the liquid (i.e. the lack of long-range order) might play a crucial role in the dynamics. We investigate this experimental result with a combination of newly developed \textit{ab~initio} techniques, and introduce a semi-classical model for HHG in liquids. Our proposed model takes electron scattering into account and successfully reproduces the observed wavelength-independence of $E_c$. We identify a key parameter in HHG from the liquid phase - the effective mean-free path ($\lambda_{\rm MFP}$) - which we extract from measurements using the extended semi-classical model. Our results shed light on fundamental strong-field-driven processes in liquids, that are distinct from what happens in either gas phase or solid-state environments, and form the basis of a first intuitive picture of HHG in liquids.

\section{Results}
\subsection{Effect of electron scattering on HHG}
One noticeable difference between HHG in dilute gases and in condensed phases is the significance of electron scattering in the latter. We therefore start our analysis by formulating a semi-classical real-space trajectory picture similar to the TSM of gas-phase HHG, but include scattering from the beginning. Within this picture, harmonic photons are emitted as a result of electron trajectories that recombine with their parent ion. We assume that an electron may be photo-excited to the conduction band of the liquid at any time, $t_\mathrm{ion}$ during the laser cycle. Following this, Newtonian equations of motion can be analytically solved to obtain the electron trajectory along the laser polarization axis ($x(t)$, given in atomic units):
\begin{equation}
x(t) = \frac{qE_0}{m\omega^2} [\cos(\omega t)-\cos(\omega t_\mathrm{ion})+\omega(t-t_\mathrm{ion})\sin(\omega t_\mathrm{ion})],
\end{equation}
where $E_0$ is the laser field's peak amplitude, $q$ and $m$ are the electron charge and mass, and $\omega$ is its angular frequency.
Recombining trajectories arise by demanding that the electron returns to its initial position at time, $t_\mathrm{rec}$, i.e. $x(t_\mathrm{rec})=0$. This generates a set of short and long electron trajectories that are initiated at $t_\mathrm{ion}$, recombine at $t_\mathrm{rec}$, and have an emission energy of $\Omega=I_p+0.5(\,dx/dt)^2$. The resulting cutoff in the absence of scattering is then obtained semi-analytically as $\Omega_\mathrm{cutoff}=I_p+3.17U_p$, where $U_p=E_0^2/4\omega^2$ is the classical ponderomotive energy. In the following, we consider only the trajectories for which the electrons return to their parent molecule. We extend this model to include scattering processes of electrons with neighboring molecules: as a simple approximation, we assume that any trajectory that exceeds a characteristic excursion length (denoted as \textit{l}$_\mathrm{max}$) will likely scatter and therefore not contribute significantly to the HHG emission (see Fig.~\ref{fig:one}A). That is, we neglect the contribution of scattered electrons to HHG. The electron could also recombine to other centers (as reported in the case of solids \cite{you17a}). However, semi-classical calculations show that this leads to a higher-energy cutoff and a different behavior under elliptically polarized light than measured experimentally (see Extended data Fig. 3), which allows us to discard this channel as the dominating one for our driving conditions. 

\begin{figure*} [!htb]
\includegraphics[width=0.8\textwidth]{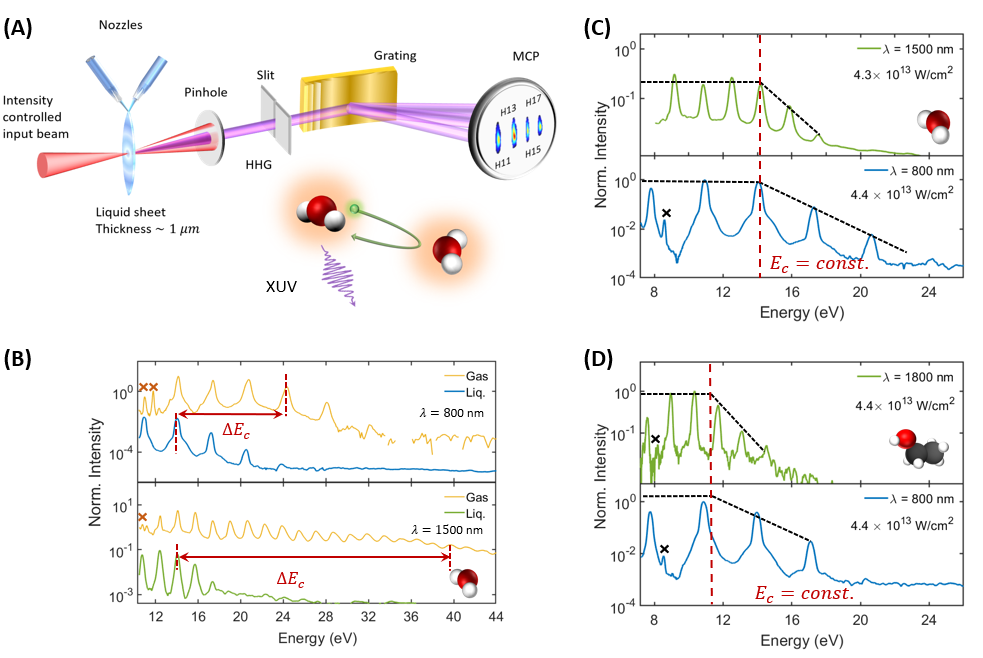}
\caption{{\bf Wavelength scaling of HHG in liquids and gases}
\textbf{(A)} 
Schematic of the experimental setup. Laser pulses with a central wavelength of 800 nm, 1500 nm or 1800 nm are focused on the flat liquid jet to generate high harmonics. The generated high harmonics pass through a slit into the XUV spectrometer that disperses and records the different harmonic orders. 
\textbf{(B)} 
High-harmonic spectra from liquid and gaseous water recorded under identical conditions using an 800-nm (top) or 1500-nm (bottom) driver. The difference in cut-off is indicated by the red arrows. The normalized liquid spectra are vertically scaled by a factor of 500 (at 800 nm) and 1000 (at 1500 nm) with respect to the gas spectra for better visualization. 
\textbf{(C)} 
High-harmonic spectra from liquid water recorded at different wavelengths, but very similar intensities.
\textbf{(D)} 
High-harmonic spectra from liquid ethanol recorded at different wavelengths, but very similar intensities.
In all panels, crosses mark the harmonics reflected in the second diffraction order of the grating.}
\label{fig:two}
\end{figure*}
This effectively translates to the following constraint for linearly polarized light - only recombining trajectories that uphold $|x(t_\mathrm{ion}<t<t_\mathrm{rec})|<l_\mathrm{max}$ emit harmonics. The set of trajectories that fulfills these equations of motion are identical to those in the gas-phase case, except that trajectories extending beyond $l_{max}$ are not included. As a consequence, this introduces a mechanism that modifies $E_c$ and its scaling, and also naturally reduces $E_c$ compared to the gas phase by $\Delta E_c$. 

Typical results for the model are presented in Fig.~\ref{fig:one}(C) where the cutoff energy follows the standard TSM prediction for short wavelengths (where both short and long trajectories do not surpass $l_\mathrm{max}$), but rapidly saturates around 800~nm where the cutoff trajectories in the gas phase exceed a distance of few angstroms. We will show below that this simple model reproduces the main features of both measurements and \textit{ab-initio} calculations. We note that the non-scaling of the cutoff with wavelength is reproduced by the semi-classical model for any choice of $l_\mathrm{max}$, which only changes the maximal $E_c$ (see Figure~\ref{fig:one}(B)). Moreover, this behavior does not depend on the laser intensity (see Supplementary Material (SM), Section S4.C and Fig.~S6). 

We emphasize that this simple picture likely does not capture the full physics of strong-field light-matter interactions in the liquid state. Nonetheless, the fact that the very peculiar cutoff behavior (compared to other phases of matter) is reproduced, is encouraging. We also note that some of the approximations utilized here might not be accurate in the liquid phase (e.g. the strong field approximation (SFA) or neglecting multi-center recombination), but: (i) corrections accounting for these effects can conceptually be added, and (ii) the characteristic physical behaviour of the cut off is independent of these approximations, as demonstrated by the \textit{ab-initio} results shown in Section C.

\subsection{Experimental results}

The experimental setup is shown in Fig.~\ref{fig:two}(A). It consists of a laser system delivering $\sim$30-40~fs laser pulses with adjustable central wavelength (800-1800~nm) and a high-vacuum chamber containing the liquid flat-jet system and a flat-field imaging spectrometer. Further details are given in the Methods section and in the SM, Section S1. We have measured high-harmonic spectra of water (\ce{H2O}) and ethanol (\ce{CH3CH2OH}) from the liquid and gas phases of each species at different wavelengths. A typical background-corrected HHG spectrum of water is shown in Fig.~\ref{fig:two}(B). The liquid- and gas-phase spectra are recorded back-to-back to minimize drifts in the experimental parameters. Figure~\ref{fig:two}(B) presents HHG spectra for \ce{H2O} with the top and bottom panels directly comparing the gas- and liquid-phase signals recorded with 800-nm and 1500-nm drivers, respectively. The liquid-phase harmonics are roughly ten times brighter than the gas-phase harmonics. Both spectra exhibit a distinct plateau, followed by a sharp cut-off region where the harmonic yield drops exponentially. Notably, the cut-off energy $E_c$ is around H9 in the liquid spectrum and H17 in the gas spectrum. To reduce uncertainty, we determine the cut-off energy following the formalism elaborated in section S3 of the supplement. In brief, as the harmonic yield in the cut-off region is expected to decay exponentially, the logarithmic value of the harmonic yield is fitted to a linear function of the harmonic energy. In contrast, the harmonic yield for the plateau harmonics remains constant as a function of energy. The intersection of these two lines (the linear fit of the log(yield) values as a function of the harmonic energy in the cut-off region and the line indicating the average log(yield) value of the plateau harmonics) defines the cut-off energy or the end of the plateau region. We find that the liquid phase shows a much-reduced cut-off compared to the gas phase. For generation in \ce{H2O}, the gas-to-liquid difference $\Delta E_c$ is about 10~eV at 800~nm and about 26~eV at 1500~nm. This observation is a first hint at different dominant mechanisms in each phase of matter, which prevents the emission of higher energy photons from the liquid.

We next explore the wavelength scaling of $E_c$. This basic property reveals information about the laser-driven electron dynamics in the liquid phase. Figure~\ref{fig:two}(C) and (D) show measured high-harmonic spectra from water and ethanol at two different wavelengths, respectively. All spectra display the characteristic envelope with a plateau and a sharp cut-off region. This allows us to define the cut-off energy $E_c$ as the intersection point of two lines that connect the plateau and cut-off intensities, respectively. The details of this procedure that is followed throughout this work are given in Section~S3. For both liquids, all spectra share the same cut-off energy of the plateau, i.e. $E_c$ = 14.2~eV and $E_c$ = 11.4~eV in the case of water and ethanol, respectively. These results substantially differ from the gas-phase results, as well as the standard TSM, which both show that for the used laser intensities the cut-off should have extended by $\sim$25~eV between the 800~nm and 1500~nm drivers. 

\begin{figure*} [!htb]
\includegraphics[width=0.8\textwidth]{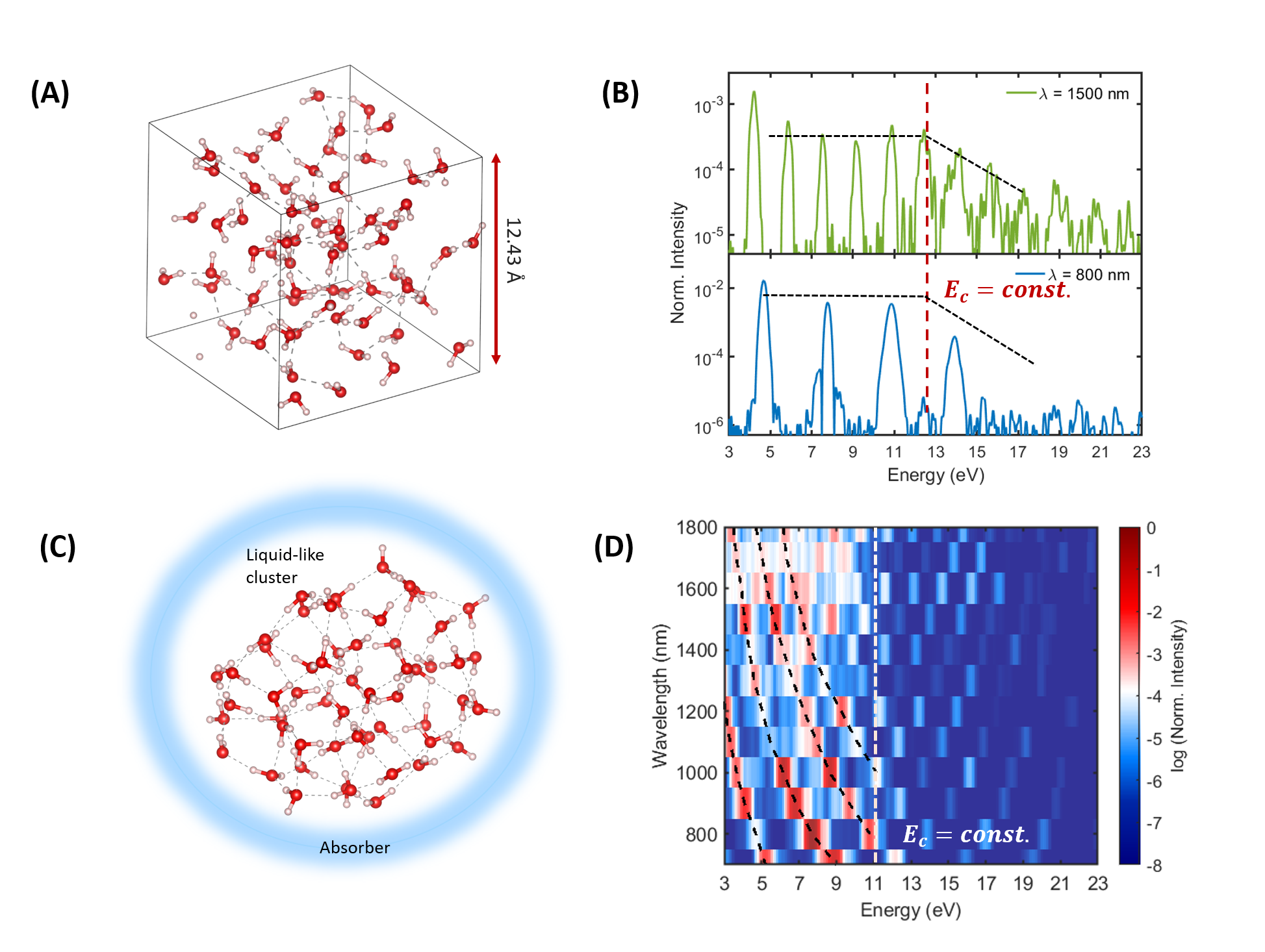}
\caption{
{\bf TDDFT calculations in the supercell approach}
\textbf{(A)}
A schematic representation of liquid water in supercell approach. There are 64 water molecules in a cubic cell at the experimental density of 1~gr/cm$^2$; periodic boundary conditions are implemented in 3D where each lattice vector is of length $\sim$ 12.43 \AA.
\textbf{(B)} 
HHG spectra calculated for liquid water using two different driving wavelengths. The peak intensity for all wavelengths is the same and it is equal to approximately 20~TW/cm$^2$. These HHG spectra are averaged over 5000-5500 water molecules in the liquid phase (for details see the SI and Ref.~\cite{Zahra2021}). 
{\bf TDDFT calculations in the cluster approach} \textbf{(C)} Illustration of the computational approach with a cluster radius of $\sim$ 15.5 \AA.
\textbf{(D)}
Wavelength scaling of high-harmonic spectra calculated for a constant peak intensity.} 

\label{fig:three}
\end{figure*}

Notably, $E_c$ in ethanol is $\sim$3~eV smaller than in water. The difference in cut-off energies between these liquids is substantially larger than the difference in their band gaps ($\sim8~$eV for \ce{H2O} and $\sim 8.5~$eV for ethanol \cite{Schreck2014,yin2015ionic}). This is a crucial point, since in the gas phase, and within the standard TSM, the cut-off should only vary by the difference of these value. The larger variation is indicative of the fact that the liquid structure, and more precisely, the electron dynamics in the liquid-phase, are playing an additional and yet to be specified role. 

A critical aspect of HHG in liquids consists in ensuring that the measured signals originate from the bulk liquid phase. This requires explicitly excluding HHG emission from the evaporating gas phase, as well as HHG from the gas-liquid interface. A complete experimental separation of HHG emission from the gas and liquid phases has been achieved by using the wedge-like geometry in the upper part of the liquid jet, as shown in Extended Data Fig. 1. Additional experiments with the liquid jet placed at an angle of 45° with respect to the driver-beam propagation allow us to exclude HHG from the liquid-gas interface to contribute a measurable signal because of the absence of any even harmonics (see Extended Data Fig. 2).

We thus reach two main conclusions: (i) in the liquid-generated high-harmonic spectra the position of the cut-off depends on the nature of the liquid sample, and (ii) the cut-off energy is wavelength independent, at least in water and ethanol. In what follows, we will show that these results are reproduced by \textit{ab-initio} calculations.

\begin{figure*} [!htb]
\includegraphics[width=0.8\textwidth]{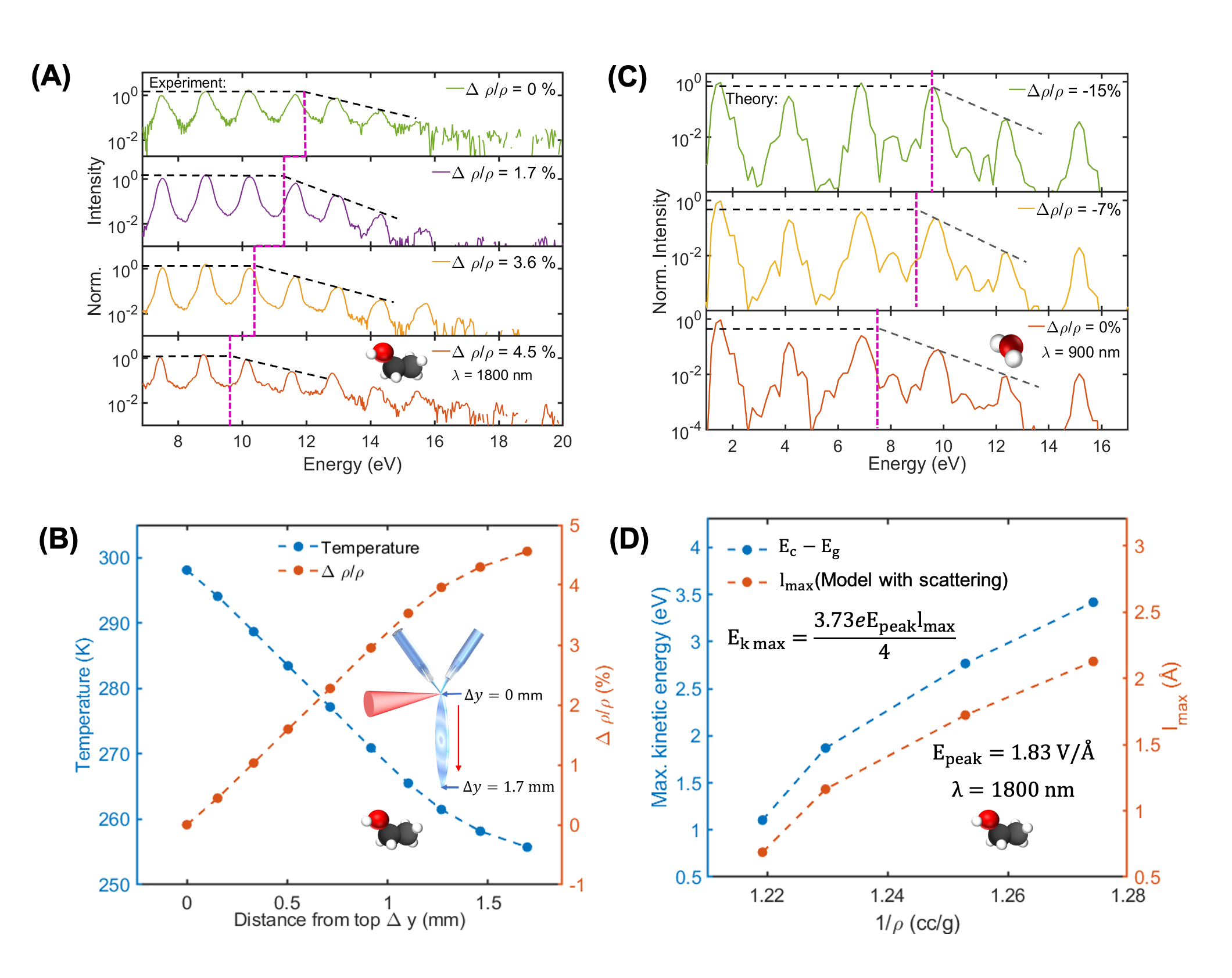}
\caption{
\textbf{Scaling of the cut off with the density of the liquid}
\textbf{(A)}
Measured high-harmonic spectra (1800 nm driver, $6\times 10^{13}$~W/cm$^2$) as a function of vertical position on the liquid flat-jet, corresponding to the indicated change in density. The change in density has been determined from an absolute temperature measurement carried out by Raman thermometry \cite{chang2022}, as shown in \textbf{(B)}.
\textbf{(C)}
Spectra calculated with the \textit{ab-initio} cluster approach (900 nm, $4 \times 10^{13}$~W/cm$^2$) for varying densities ($\Delta\rho/\rho$ = -15$\%$, -7$\%$, 0$\%$, 7$\%$). The harmonics above the cut-off show a systematic decrease in yield with increasing density.
\textbf{(D)} Maximal observed kinetic energy as a function of the inverse density for ethanol and the corresponding $l_{\rm max}$ obtained from the formula shown in the inset (derived in SM, Section S4.D). The vacuum electric field is corrected using the 1800-nm refractive index in ethanol, to obtain the peak electric field (E$_{peak}$) inside the liquid.
}
\label{fig:four}
\end{figure*}

\subsection{Numerical results: time-dependent density-functional theory}

We now compare these experimental findings and the results of our simple model to two newly developed \textit{ab~initio} techniques for describing the strong light-matter response of liquids. Figure~\ref{fig:three} presents simulated HHG spectra from liquid water that are based on a combination of well-established Car-Parrinello molecular dynamics (CPMD) \cite{car1985unified} and time-dependent density-functional theory (TDDFT) \cite{runge1984} simulations in a periodic supercell including 64 water molecules at the experimental density of 1~g/cm$^3$ and temperature of 300~K (for details see Ref. \cite{Zahra2021} and the SM, Section S4.A). This is the first realistic, currently tractable description of HHG in liquids. TDDFT naturally includes mean-free path effects as it includes electron-electron and electron-ion scattering. Overall, very good agreement with the experimental results is observed, and most importantly, the cut-off energy and its wavelength independence are well reproduced in Figure~\ref{fig:four}. Moreover, a time-frequency analysis of the TDDFT results (see Extended Data Fig. 4) shows that only very short  electron trajectories contribute to the HHG spectra, in agreement with our semi-classical model.
Note that since the DFT-GGA underestimates the liquid-water band gap, the calculated HHG cutoff is about 1.5~eV lower than the experimental value. This numerical approach qualitatively reproduces the experimentally observed weak dependence of the cut-off on the laser intensity (see discussion in SM, Section S4.A). This further confirms that the above experimental findings are a signature of the microscopic mechanism in the liquid phase and not the result of macroscopic effects, which are absent in our theoretical modeling.
\begin{figure} [!htb]
\includegraphics[width=0.5\textwidth]{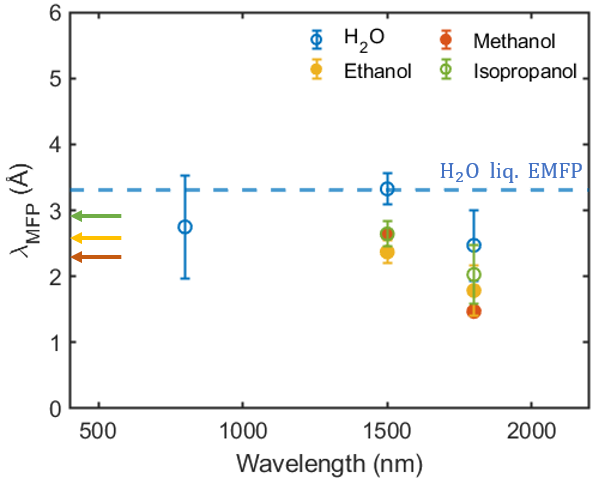}
\caption{
\textbf{Comparison of the determined electron mean-free paths with literature data.}
Mean-free paths for electron scattering in the liquid phase determined from the experimentally observed cut-off energies ($E_c$) as a function of laser wavelength. The dashed blue line indicates the elastic mean-free path for liquid water \cite{schild20a} at $E_k = 4.5$~eV. The arrows indicate the value of the mean-free paths obtained from the integral elastic scattering cross sections of the corresponding alcohols in the gas phase \cite{khakoo08a,bettega11a} at $E_k$ of $\sim$ 3-4.5~eV, using the number densities obtained from the densities of different liquids at 20$^\circ$C \cite{liqdense}.
}
\label{fig:five}
\end{figure}

This result is complemented by a second set of \textit{ab-initio} TDDFT calculations based on molecular clusters which employ some additional approximations (for details, see ref. \cite{Ofer2021}). The advantage of this approach is that it allows for faster calculations while still leading to accurate results; thus, it can be employed for a more detailed numerical study. Figure~\ref{fig:three}(B) shows simulation results of HHG in liquid water with the cluster approach for many wavelengths at a fixed laser intensity. Clearly, the same trend is observed and the cut-off is independent of the wavelength, at least in the range of 500~nm - 1800~nm. In the SM, Section S4.B, we show that the cut-off with the cluster approach is similarly weakly dependent on the laser intensity (above a saturation intensity of $5 \cdot 10^{13}$~W/cm$^2$), and that the wavelength independence of the cut-off is maintained for other laser intensities, as well. With the cluster approach, we also performed calculations for two additional liquids (ammonia, \ce{NH3}, which is polar, and methane, \ce{CH4}, which is non-polar). In the SM, Section S4.B, we show that HHG calculations in liquid \ce{NH3} and liquid \ce{CH4} also predict the same wavelength independence of $E_c$. These results, in combination with our measurements, lead to the conclusion that this characteristic non-scaling of $E_c$ is a fundamental general and unique property of the liquid-phase HHG, and applies both for polar and non-polar liquids. These accurate quantum-dynamical simulations reproduce and complement our experimental findings, which validates the broad applicability of our conclusions. 

\section{Discussion and conclusions}

We have so far demonstrated the wavelength-independence of $E_c$, both experimentally and theoretically, (Figs.~2 and 3) and we have shown that a scattering-limited trajectory model reproduces this behavior (Fig.~1). Importantly, a main conclusion arising from our results is that if $E_c$ is limited by the electron mean free path $\lambda_{\rm MFP}$ in the liquid, then $E_c$ should scale with the density of the liquid. Figure 4 demonstrates that this is the case, both experimentally (A) and theoretically (C).

Very recently, some of us have reported the first measurements of the temperature of liquid flat-jets \cite{chang2022}. The temperatures, measured by Raman thermometry under conditions identical to those of the present HHG experiments, range from $\sim$300~K at the top, to $\sim$255~K at the bottom of the first sheet, translating to a density variation of close to 5$\%$ (Fig.~4(B)) in the case of ethanol \cite{liqdense}. Over this range of conditions, $E_c$ is found to decrease by $\sim$2~eV (panel (A)). The same trend is also observed in the calculations performed on liquid water (panel (B)). 
Having measured $E_c$ over a range of densities, we can now verify how the maximal energy $E_{k,{\rm max}}$, gained by the electron from the driving laser field, scales with the density. Experimentally, we use $E_{k,{\rm max}}=E_c-E_g$, where $E_g$ is the band gap of the liquid. We find that $E_{k,{\rm max}}$ scales linearly with the inverse density (blue symbols in Fig.~4(D)). This type of scaling precisely corresponds to the prediction of our simple trajectory-limited model (Fig.~1), because $\lambda_{\rm MFP}$=$1/(n\sigma)\propto 1/(\rho\sigma)$. 
This conclusion is further supported by converting the measured $E_c$ to the corresponding maximal excursion length $l_{\rm max}$. As we show in the SM, Section S4.D, we find that $E_{k,{\rm max}}=3.73/4*e*E*l_{\rm max}$. A direct consequence of this relation is that it allows us to retrieve $l_{\rm max}$ from the experimental spectra, provided that they were recorded under conditions where the wavelength-independence of the cut-off is observed, which is the case here (see Fig.~2D). The orange symbols in Fig.~4D show that $l_{\rm max}$ also scales linearly with the inverse density.
We therefore conclude that all experimental and theoretical evidence available at present agrees on the fact that $E_c$ is proportional to the maximal excursion length of the laser-driven electrons in the liquid phase. This suggests that it should be possible to accurately determine effective electron mean-free paths (MFPs) from liquid-phase HHG spectra.
Electron MFPs play a very important role in describing electron-driven processes in the liquid phase, but they are notoriously difficult to measure and calculate, at least at low energies. The interest in developing new methods for accessing these quantities is therefore considerable and relevant for many physical processes. Here, we do not attempt to determine the MFPs with high precision because this would require a more sophisticated scattering model, including a large number of different scattering channels (see Ref. \cite{gadeyne22a} and references therein). Instead, we aim at retrieving an effective MFP, which is best thought of as containing all types of scattering processes. Since the elastic scattering cross sections are by far dominant at the very low kinetic energies ($\sim$10 eV) of interest here \cite{song21a,thuermer13a,schild20a,gadeyne22a}, we compare our results to the elastic MFPs in Fig.~\ref{fig:five}. In this comparison, we use $\lambda_{\rm MFP}=l_{\rm max}$, taking into account that the electron travels up to the maximal excursion length before getting scattered. In the SM (Fig.~S10), we show that this simple approximation is physically meaningful because replacing the ''sharp'' truncation of the trajectories (at the travel distance $l_{\rm max}$) with an exponential distribution of path lengths (inherent to the definition of $\lambda_{\rm MFP}$) leaves $E_c$ unchanged.
Figure~\ref{fig:five} compares the $\lambda_{\rm MFP}$ values obtained from the HHG spectra (symbols) with the available literature values. In the case of liquid water, we are comparing to the most recent MFPs (blue dashed line), which were determined from a Monte-Carlo simulation of experimental liquid-microjet data using the most accurate {\it ab-initio} differential scattering cross sections available to date \cite{schild20a}. In the case of the alcohols, liquid-phase MFPs have to our knowledge not been reported in the literature so far. We are therefore comparing our results to MFPs determined from the corresponding experimental gas-phase elastic scattering cross sections and the known number densities of the alcohols. The agreement is very good in all cases, confirming the possibility to retrieve electron MFPs from liquid-phase HHS. 
To summarize, we explored here the microscopic mechanisms responsible for liquid HHG with a combination of experimental and theoretical methods. Our measurements in water and ethanol show that contrary to crystals and gases, the cut-off energy in liquid HHG is mostly independent of the laser wavelength. Microscopic quantum mechanical calculations based on both supercells and clusters agree with this result, and show that it extends to other liquids and laser conditions. We proposed an extended semi-classical model for the electron dynamics in the liquid to explain this result. The model incorporates effects of ultrafast scattering of electrons off neighboring molecules, which are shown to reduce the HHG cut-off compared to the gas phase case through the spatial limitation of electron trajectories. The model reproduces well the wavelength independence of the HHG cut-off and highlights the importance of the electron mean-free path in liquids, indicating that this quantity is imprinted onto the high-harmonic spectra and can be retrieved. We also expect that our results would be highly relevant for HHG from amorphous solids \cite{luu18b,you17b}. Our work paves the way to a deeper understanding of the strong-field dynamics in disordered condensed phases, and to resolving attosecond dynamics in liquids.

\section*{Methods}
The experimental set-up consists a 1 kHz Ti:Sapphire laser delivering $\sim$ 30~fs pulses at 800~nm. Driving wavelengths of 1500~nm and 1800~nm are obtained by optical parametric amplification (OPA) of the 800~nm pulses, respectively. The driver beams are focused with a spherical mirror on a sub-micron-thin liquid flat-jet target, further described in Refs. \cite{Luu2018,yin2020}. The beam intensities are calculated from the harmonic cut-off energy of the gas-phase measurements, using the semi-classical TSM\cite{Corkum1993}.
The emerging high harmonics are analysed with a custom-built XUV spectrometer consisting of an aberration-free flat-field grating (SHIMADZU) and a multi-channel plate (MCP) coupled with a phosphor screen. The phosphor-screen image is recorded by a CCD camera. Each spectrum is typically integrated for 10 - 20 ms and measured 20 times. These spectra are averaged prior to all subsequent analysis. For the ellipticity dependent studies, an elliptically polarized 800 nm light wave generated using a combination of a rotating half-wave plate(HWP) and a fixed quarter-wave plate (QWP). The rotation of the HWP axis from 0$^{\circ}$ to 22.5$^{\circ}$ with respect to the QWP axis, changes the polarization of the input light from linear to circular keeping the axes of the polarization ellipse fixed. At a number of different ellipticities, ranging from $\epsilon$ = 0 for linear polarization to $\epsilon$ = 1, for circular polarization, the harmonic spectrum was measured for different liquids. Further details of the experimental and theoretical methods are described in the SM, Section S1 and the extended data Figure~1.

\section*{Data Availability}
The datasets generated during and/or analysed during the current study are available from the corresponding author on reasonable request.

\section*{Code Availability}
The Octopus package used for the TDDFT calculations is publicly available. The remaining computer codes are available from the corresponding author on reasonable request.

\begin{acknowledgments}
The authors thank Andreas Schneider and Mario Seiler for their contributions to the construction and improvements of the experiment and Y.P. Chang, T. Balciunas and T.T. Luu for scientific discussions.
We acknowledge financial support from ETH Z\"{u}rich and the Swiss National Science Foundation through grant 200021-172946. 
This work is supported by the Deutsche Forschungsgemeinschaft (DFG) through the priority program QUTIF (SOLSTICE-281310551) and the Cluster of Excellence `CUI: Advanced Imaging of Matter'- EXC 2056 - project ID 390715994, Grupos Consolidados (IT1249-19), and the Max Planck - New York City Center for Non-Equilibrium Quantum Phenomena. The Flatiron Institute is a division of the Simons Foundation.
AM acknowledges the support of the InterMUST-AoW PostDoc Fellowship. ZY acknowledges financial support from an ETH Career Seed Grant No SEED-12 19-1/1-004952-00. ON acknowledges support from the Alexander von Humboldt foundation and a Schmidt Science Fellowship.
\end{acknowledgments}

\section*{Author Contributions}
A.M., Z.Y. and V.S. performed the experimental measurements and data analysis. 
O.N. carried out the cluster calculations and the numerical extended semiclassical model for HHG in liquids.
Z. N. carried out the supercell calculations.
N. T.-D. developed the analytical analysis of the HHG cutoff from the extended semiclassical model.
A.R. and H.J.W. supervised the work.
All the authors participated in the discussion of the results and contributed to the manuscript.

\section*{Competing interests}
The authors declare no competing interests.

\bibliography{ref.bib,attobib.bib}

\setcounter{figure}{0} 
\begin{figure*} [!htb]
\includegraphics[width=0.9\textwidth]{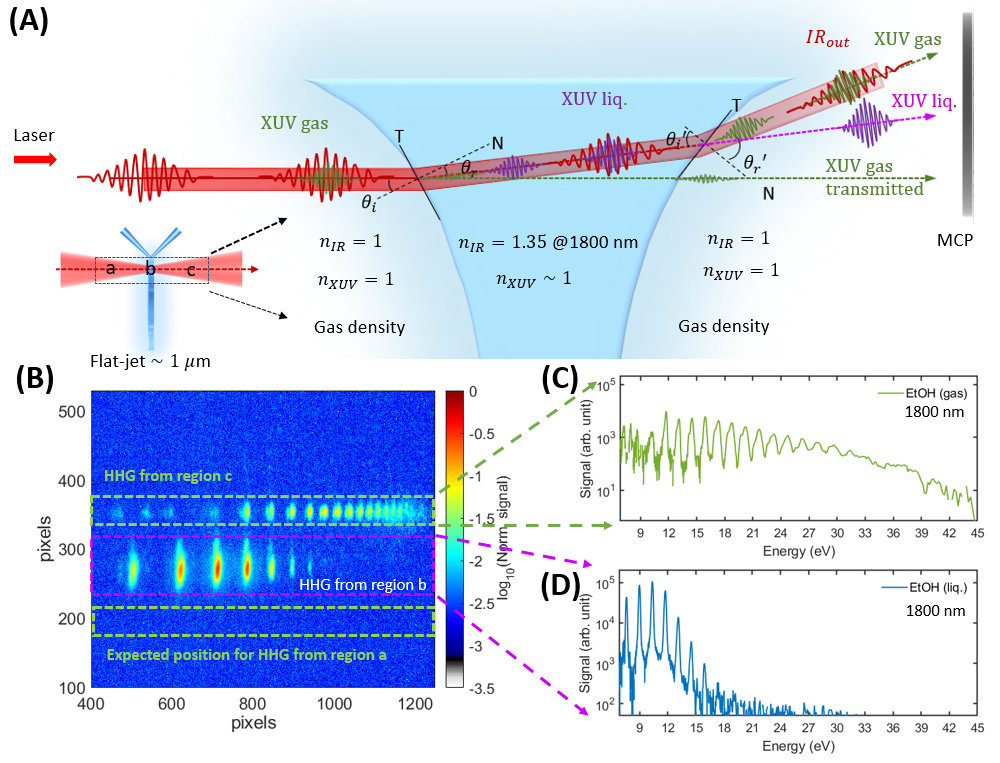}
\renewcommand{\thefigure}{Extended data \arabic{figure}}
\caption{\textbf{(A)} Schematic of the IR beam optical path and the high-harmonic radiation in the top part of the liquid jet (see also Ref. \cite{Luu2018}). The bottom-left inset of panel (A) shows the overview with regions labeled a-c. The high-harmonic emission from the gas before the jet (region a) is absorbed in the first few layers of the liquid ($\sim$10 nm\cite{hayashi15a}). Any XUV radiation generated in front of the liquid jet passes unrefracted through the liquid medium. This is because the index of refraction of liquid ethanol (as well as other alcohols and water) in the relevant XUV range (7-44 eV) is very close to 1 \cite{hayashi15a}. The high-harmonic radiation generated in the bulk liquid (region b) is not refracted significantly at the exit liquid-gas interface. As a result, the XUV beam is refracted negligibly in comparison to the IR beam (the refractive index $n_{IR}$ is 1.35 at 1800 nm in ethanol \cite{SANI2016137}), as it exits the liquid phase into region c. In the schematic $\theta_i$ and $\theta_i'$ denote the angle of incidence of the IR beam at the entry and exit surfaces of the liquid jet. $\theta_r$ and $\theta_r'$  denote the angle of refraction of the IR beam at the entry and exit surfaces of the liquid jet and the lines T and N denote the tangent and normal to each surface. The strong refraction of the IR beam at the exit liquid-gas interface gives rise to the spatial separation of the liquid-phase harmonics (generated from the bulk liquid (region b)) and the gas-phase harmonics (generated from after the liquid jet (region c)) of ethanol at 1800 nm, on the micro-channel plate (MCP) detector. The green dashed box on top of the liquid-phase harmonics indicates the region of interest on the detector selected for extracting the gas-phase harmonic spectrum \textbf{(C)}. Similarly, the purple dashed box indicates the region of interest selected for extracting the liquid-phase harmonic spectrum \textbf{(D)}. The green dashed box below the liquid-phase harmonics indicates the region where the transmitted gas XUV (generated in front of the liquid jet (a)) is expected to be incident on the detector. The absence of these harmonics is explained by their expected absorption in the liquid jet.}

\label{fig:ED1}
\end{figure*}

\begin{figure*} [!htb]
\includegraphics[width=0.8\textwidth]{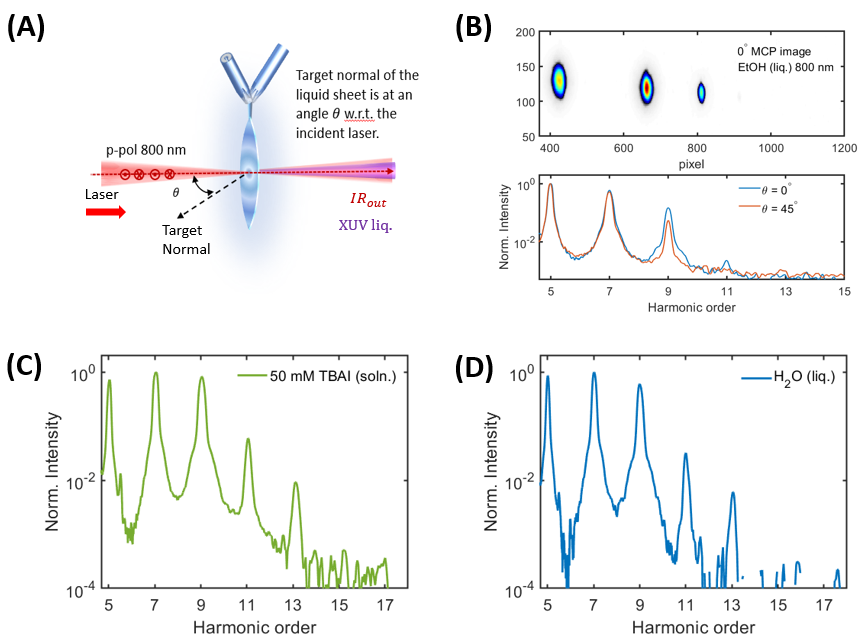}
\renewcommand{\thefigure}{Extended data \arabic{figure}}
\caption{\textbf{(A)} Schematic showing the flat-jet orientation with respect to the incident laser direction used for excluding the presence of surface/interface-generated harmonics. The p-polarized (perpendicular to the plane of the figure) 800-nm laser beam is made incident on the flat-jet. The flat-jet is rotated such that its target normal is oriented at an angle $\theta$ with respect to the incident laser direction. If harmonics are generated at the interface where the beam exits the flat-jet, the liquid medium would break the symmetry of the light field (and therefore the electron trajectories) for the negative and positive half cycle of the electric field for any $\theta > 0$, which would lead to the generation of even harmonics. \textbf{(B)} The raw MCP image for  $\theta = 0^{o}$ (top panel) and the comparison of the harmonic spectra (bottom panel) taken for  $\theta = 0^{o}$(blue line) and  $\theta = 45^{o}$(orange line), for 800 nm laser beam incident on an ethanol flat-jet. No even harmonics are detected for the tilted jet geometry, which suggests that the detected signals are dominated by the bulk liquid with negligible interface contributions. 
\textbf{(C)} We also studied the harmonic emission from 50-mmol Tetrabutylammonium iodide (TBAI) solution at 45° jet orientation. TBAI was chosen here because of its high surface affinity caused by the hydrophobic interactions of the TBA$^+$ cation and the large polarizability of the I$^-$ anion \cite{TBAI2004}. This should further enhance the visibility of any even surface/interface harmonics arising from the breaking of the inversion symmetry of the p-polarizer driver by the liquid jet. We also compared this spectrum with that of pure water (D) for the same 45° jet orientation. However, even harmonics were not observed in any of these cases, indicating that the surface/interface contribution to the HHG response is negligible as compared to the bulk of the liquid.}

\label{fig:ED2}
\end{figure*}

\begin{figure*} [!htb]
\includegraphics[width=\textwidth]{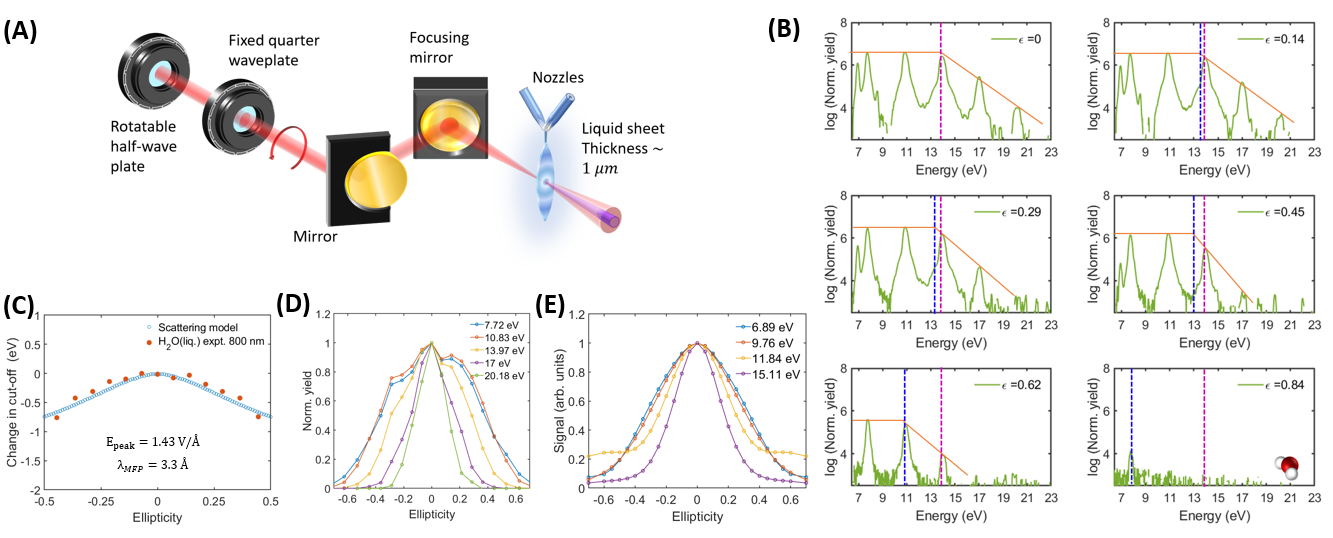}
\renewcommand{\thefigure}{Extended data \arabic{figure}}
\caption{\textbf{(A)} Experimental setup for studying the ellipticity dependence of high-harmonic generation from liquids. The rotating half-wave plate and fixed quarter-wave plate geometry is used to keep the axes of the elliptically polarized driving field fixed as the light changes from linear to circular polarization. \textbf{(B)} Harmonic spectra obtained from liquid water at 800 nm for different driver ellipticities. The pink dashed line in each sub-figure represents the cut-off energy for linearly polarized light. The blue dashed line indicates the actual cut-off for each of these spectra. It is observed that as the light changes towards circular polarization ($\epsilon$=1), the cut-off shifts to lower energies. \textbf{(C)} Comparison of the experimentally determined change in cut-off and that calculated from the MFP-limited scattering model which considers only recombination with the parent molecule. Details of the theoretical scattering  model are given in the supplementary section S10. \textbf{(D)} experimental data showing high-harmonic yields vs. ellipticity. \textbf{(E)} Data from cluster calculations showing the high-harmonic yields vs. ellipticity. These calculations were performed using a 900-nm driving field with a peak intensity of 4$\times$10$^{13}$ W/cm$^2$. }
\label{fig:ED3}
\end{figure*}

\begin{figure*} [!htb]
\includegraphics[width=0.9\textwidth]{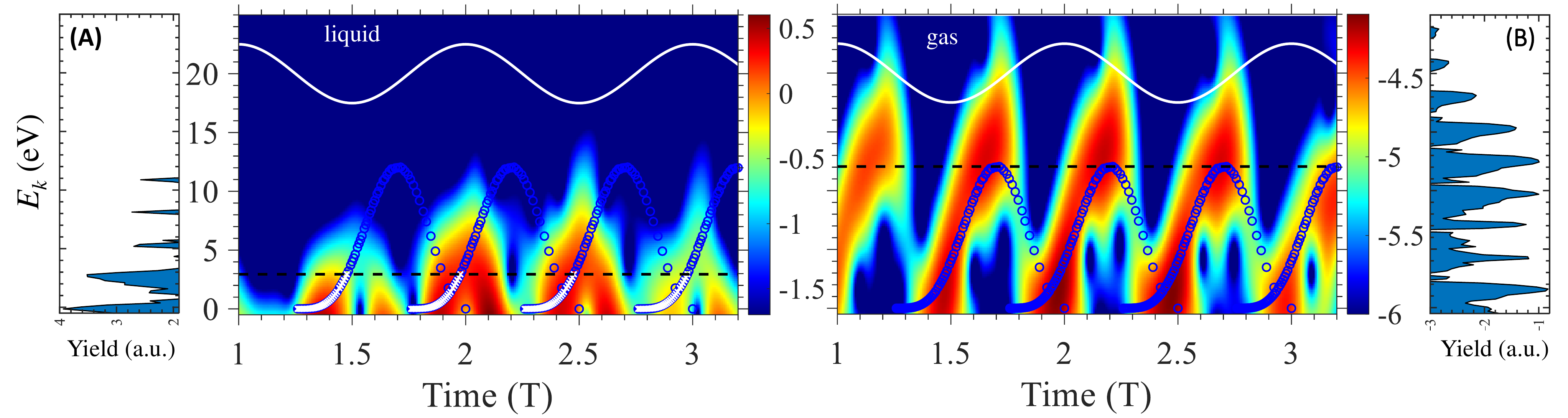}
\renewcommand{\thefigure}{Extended data \arabic{figure}}
\caption{\textbf{(A)} Time-frequency analysis for calculated HHG emission from the liquid cluster, comparing the ab-initio results to the simple trajectory-based model that includes MFP-limitted trajectories. The y-axis shows the emitted HHG photon energy with respect to the HOMO-LUMO gap in the cluster, which corresponds to the excess kinetic energy of the electrons upon recombination. The x-axis is given in units of an optical cycle. The gas-phase expected SFA trajectories are shown in blue circles, while the remaining short trajectories that do not exceed the MFP are highlighted with white crosses. The corresponding HHG spectra are plotted for reference, along with the laser electric field (in solid white). The cutoff from the semi-classical model is indicated in dashed black. (B) Same as (A) but for an isolated molecule, where the response is orientation averaged to describe an un-oriented gas of water molecules. The cutoff in dashed black indicates the one obtained from SFA trajectories, and only SFA trajectories from the HOMO orbital are plotted (while some low yield higher energy emission still exists due to emission from the HOMO-1 and HOMO-2 orbitals). Both plots show the same kinetic energy scale and are calculated for 900-nm laser driving with an intensity of 5$\times$10$^{13}$ W/cm$^2$. The plot clearly indicates that in the liquid phase there is a dominant very-short trajectory contribution, whereas longer trajectories are strongly suppressed. The ab-initio cutoff for the liquid agrees very well with the prediction from the semi-classical model that incorporates the MFP as extracted from experiments. The time-frequency structure for the gas-phase similarly agrees with the expected structure, validating the approach. We note an additional emission contribution in the liquid phase at longer times that possibly corresponds to some other scattering process or a different emission channel that will be investigated in future work. The time-frequency analysis is obtained with a Gabor transform with a Gaussian window of T/3, and the isolated molecule calculations are obtained with the methodology described in ref. \cite{Ofer2021}.}

\label{fig:ED4}
\end{figure*}

\end{document}


\preprint{APS/123-QED}

\title{Supporting Information for \\
Probing low-energy electron-scattering dynamics in liquids \\ with high-harmonic spectroscopy}

\author{Angana Mondal}
\thanks{These five authors contributed equally}
\affiliation{Laboratorium für Physikalische Chemie, ETH Zürich, Zurich, Switzerland}

\author{Ofer Neufeld}
\thanks{These five authors contributed equally}
\affiliation{Center for Free-Electron Laser Science CFEL, Deutsches Elektronen-Synchrotron DESY, Notkestra\ss e 85, 22607 Hamburg, Germany}
\affiliation{Max Planck Institute for the Structure and Dynamics of Matter, Luruper Chaussee 149, 22761 Hamburg, Germany}

\author{Zhong Yin}
\thanks{These five authors contributed equally}
\affiliation{Laboratorium für Physikalische Chemie, ETH Zürich, Zurich, Switzerland}

\author{Zahra Nourbakhsh}
\thanks{These five authors contributed equally}
\affiliation{Center for Free-Electron Laser Science CFEL, Deutsches Elektronen-Synchrotron DESY, Notkestra\ss e 85, 22607 Hamburg, Germany}
\affiliation{Max Planck Institute for the Structure and Dynamics of Matter, Luruper Chaussee 149, 22761 Hamburg, Germany}

\author{V\'it Svoboda}
\thanks{These five authors contributed equally}
\affiliation{Laboratorium für Physikalische Chemie, ETH Zürich, Zurich, Switzerland}

\author{Angel Rubio}
\affiliation{Center for Free-Electron Laser Science CFEL, Deutsches Elektronen-Synchrotron DESY, Notkestra\ss e 85, 22607 Hamburg, Germany}
\affiliation{Physics Department, University of Hamburg, Luruper Chaussee 149, 22761 Hamburg, Germany}
\affiliation{Max Planck Institute for the Structure and Dynamics of Matter, Luruper Chaussee 149, 22761 Hamburg, Germany}
\affiliation{The Hamburg Centre for Ultrafast Imaging, Luruper Chaussee 149, 22761 Hamburg, Germany}
\affiliation{Center for Computational Quantum Physics (CCQ), The Flatiron Institute, 162 Fifth Avenue, New York NY 10010, USA}

\author{Nicolas Tancogne-Dejean}
\affiliation{Center for Free-Electron Laser Science CFEL, Deutsches Elektronen-Synchrotron DESY, Notkestra\ss e 85, 22607 Hamburg, Germany}
\affiliation{Max Planck Institute for the Structure and Dynamics of Matter, Luruper Chaussee 149, 22761 Hamburg, Germany}

\author{Hans Jakob W\"orner}
\affiliation{Laboratorium für Physikalische Chemie, ETH Zürich, Zurich, Switzerland}

\maketitle
\twocolumngrid

\section{Experimental details}

\subsection{Optical beamline}
The experiments were performed with a 1~kHz regeneratively-amplified Ti:Sapphire laser (Coherent). From the laser output, 1.2~mJ beam was split and directed to a dedicated grating compressor delivering about 960~$\mu$J, $\sim$30~fs pulses with a central wavelength of 800~nm. 
The linear and vertically (s) polarized driving beam was focused on a liquid flat-jet target with a pair of metallic mirrors (protected Ag) in a Z-shaped configuration, where the first mirror was flat and the second mirror was spherical with R = 800~mm. The whole set was mounted on a common base plate on top of a manual translational stage to allow fine tuning of the overlap between the laser focus and the flat jet.

For experiments with mid-IR wavelengths (1500~nm and 1800~nm), a commercial optical parametric amplifier (OPA, HE-TOPAS from Light Conversion) pumped with 6.5~mJ, 30~fs pulses at 800~nm was used. The signal or idler beam was separated from the residual wavelengths exiting the OPA using a pair of dielectric mirrors for the particular wavelength and focused with the same set of two metallic mirrors.

\begin{figure*} [!htbp]
\includegraphics[width=\textwidth]{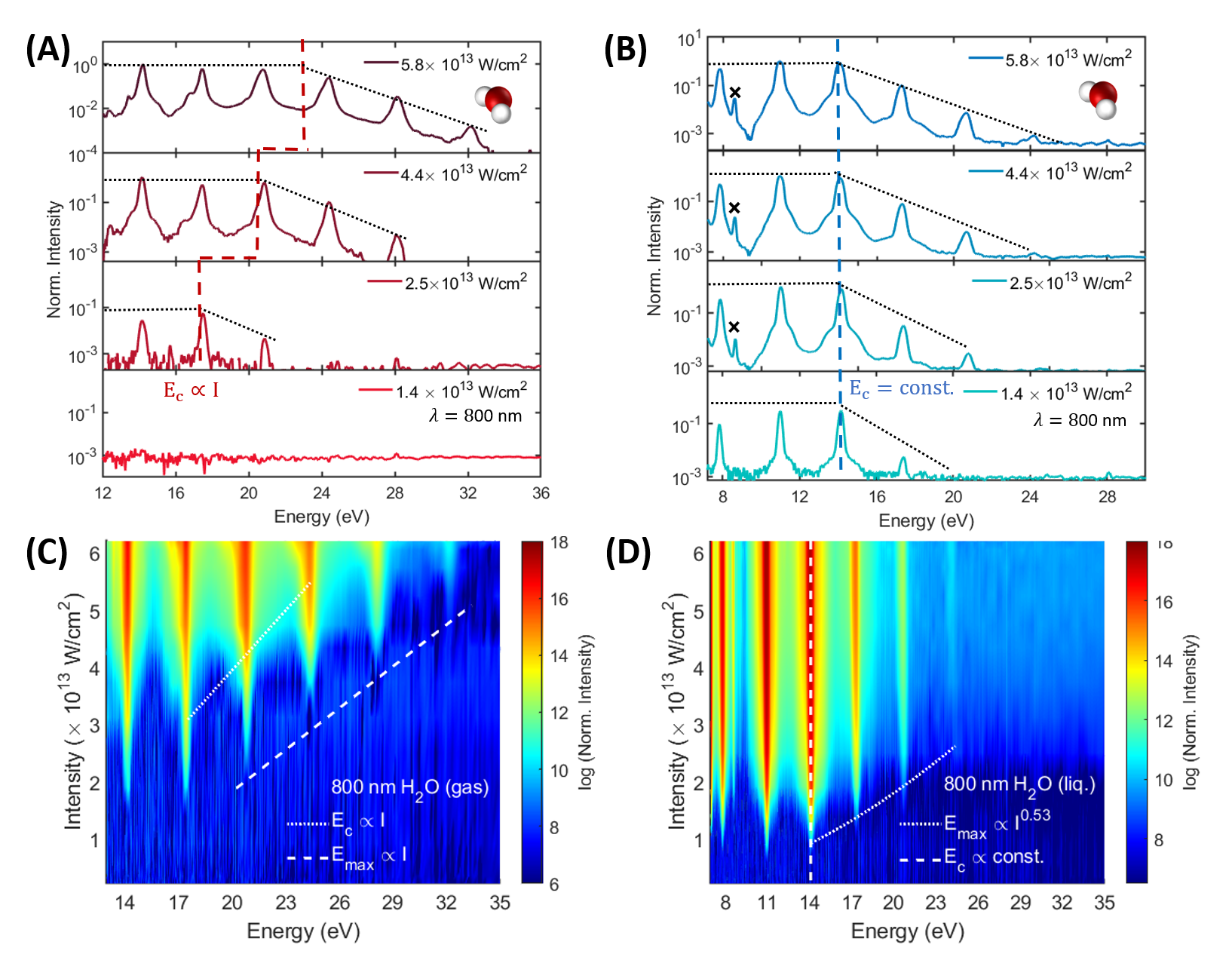}
\caption{For 800~nm driver, a comparisons between gas \textbf{(A)} and liquid \textbf{(B)} phase high-harmonic spectra of \ce{H2O} is given for three intensities. As expected, the gas phase cut-off energy depends on intensity (red dotted line) but the liquid phase is intensity independent (blue dotted line). For 800 nm driver, a comparison between gas\textbf{(A)} and liquid \textbf{(B)}  phase high-harmonic spectra of H2O is given for a range of intensities.
}
\label{fig:S1}
\end{figure*}

\subsection{Flat Jet}
The flat jet was created by two colliding cylindrical jets with about $\sim$54~$\mu$m inner diameter. The samples were pumped via an HPLC pump (JASCO) into the interaction region with a flow rate ranging from 3~ml/min to 5~ml/min. The established flat jet was about 1~$\mu$m thick \cite{yin2020}. Due to evaporation, the liquid jet was surrounded by a region with high gas density, as compared to the backing pressure in the target chamber ($\sim$$5\cdot10^{-4}$~mbar), allowing a simultaneous probing of liquid and gas phases. All liquid samples have been used without further purification: pure Milli~Q water with an electric resistivity of 18~M$\Omega$ and 99.8~\% pure alcohols (ethanol, methanol, iso-propanol) from Sigma Aldrich. 

\section{Intensity scaling of gas- and liquid-phase HHG spectra}
Figure~\ref{fig:S1} presents experimentally measured HHG spectra emitted from liquid water at 800~nm driving wavelength for different laser intensities in the range of $1.4\cdot 10^{13}$~W/cm$^2$ to $5.8\cdot 10^{13}$~W/cm$^2$. At each intensity, we compare the gas- (Fig.~S1A) and liquid- (Fig.~S1B) phase spectra. For the gas phase there is a systematic linear increase of the cutoff with laser intensity (red dashed line), in accordance with the standard TSM. In contrast, the liquid HHG cut-off energy remains roughly constant (blue dashed line). This effectively demonstrates that in the given intensity range, the cut-off energy of the liquid spectra is independent (or weakly dependent) on the peak laser intensity. We have also verified that this is the case for liquid ethanol. 

\subsection{Scaling of the highest harmonic order vs. scaling of the plateau's cutoff}
In relating the present results to the previous literature, it is important to distinguish the scaling of the highest emitted harmonic order ($E_{\rm max}$) from the scaling of the plateau's cutoff (denoted $E_c$, and discussed in the main text). Previous literature on condensed-phase HHG has mainly studied and discussed the scaling of $E_{\rm max}$. Most prominently, this is the case for Ghimire et al. for the solid state \cite{Ghimire2011} and for Luu et al. for the liquid phase \cite{Luu2018}. In contrast, the overwhelming majority of the literature on gas-phase HHG has studied and discussed the scaling of $E_c$, influenced by the definition provided by Lewenstein et al. \cite{Lewenstein1994}. Here, we show that the scaling of these two quantities is actually different in the liquid phase and thereby show that the present conclusion regarding the scalings of $E_c$ are consistent, or a least in no contradiction, with previous literature, in particular with Luu et al. \cite{Luu2018}.

Figure~\ref{fig:S1}C and D show the data in a form that highlights the scalings of both $E_{\rm max}$ and $E_c$. Panel C shows that, in the gas phase, both $E_{\rm max}$ and $E_c$ scale linearly with the intensity. Panel D, in contrast, shows that the scalings of the two quantities are noticeably different. A non-linear least-squares fitting of $E_{\rm max}$ to the functional form $E_{\rm max}\propto I^y$ returns $y=0.53$, consistent with Luu et al. \cite{Luu2018}. The independence of $E_c$ on $I$ is visible, both in Fig.~\ref{fig:S1}B and in Fig.~\ref{fig:S1}D. This validates the conclusions reached in the main text and clarifies their relation with previous work.

We also performed ab-initio calculations to test the intensity dependence of the cutoff with both the supercell (Fig.~\ref{fig:supercell}) and the cluster (Fig.~\ref{fig:clusters}) approach for liquid HHG. Both types of calculations show that the cutoff is independent of the laser driving intensity, beyond a certain threshold. In each case the cutoff increases with the laser intensity until reaching a saturation point, where the increase stops. For the supercell approach, this saturation is found at $\sim 0.25\cdot 10^{14}$~W/cm$^2$, while for the cluster approach it is observed at $\sim 0.5\cdot 10^{14}$~W/cm$^2$. Variations between the two methods are a result of slightly different electronic structures (band gaps, in particular) obtained with the two approaches. Notably, the ranges of laser intensities for which the cutoff is intensity-independent correspond well with the experimental measurements. 

Lastly, we also point out that the observed weak dependence of the HHG cutoff with respect to the laser intensity is also captured by our proposed extended semi-classical model. Figure~\ref{fig:tsm} presents the calculated HHG cutoff vs. the peak laser intensity with the extended semi-classical model, assuming 800~nm driving, and inter-molecular distances similar to those in liquid water. Initially, the HHG cutoff increases quadratically with the laser intensity just as in the standard TSM (because the trajectories are very short and do not extend beyond $l_{max}$). However, this dependence is reduced to a weak linear scaling in the rage of intensities $> 0.5 \cdot 10^{14}$~W/cm$^2$. In fact, over the intensity range of $0.5 \cdot 10^{14}$~W/cm$^2$ to ~ $10^{14}$~W/cm$^2$, the HHG cutoff only increases by $\sim 2$~eV. This result substantially differs from the gas phase, where the standard TSM predicts that the cutoff should increase by $\sim9$~eV in that region. Notably, a change of $\sim 2$~eV in the cutoff energy at 800~nm driving would only move the HHG cutoff by approximately one harmonic order, which might be difficult to detect experimentally. Thus, we overall conclude that in our examined conditions, the liquid-HHG cutoff is weakly dependent on the laser driving intensity, an effect that is described remarkably well by our suggested semi-classical picture that includes scattering. We note that potential improvements to our extended semi-classical model (e.g. relaxing some of the utilized approximations) might also improve its correspondence to the measured and ab-initio calculated results.

\section{Determination of harmonic cut-off energy from experimental data}
\begin{figure}[!h]
\includegraphics[width=\columnwidth]{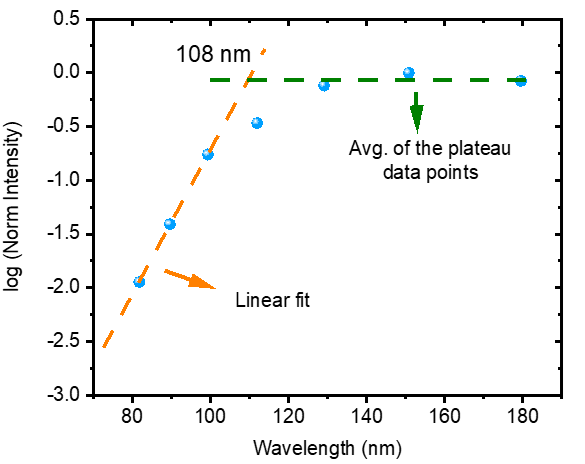}
\caption{Plot of harmonic signal as a function of harmonic wavelength. The cut-off is determined by the intersection of the green and orange lines.}
\label{fig:S6}
\end{figure}
In order to accurately determine the cut-off from the experimental spectra we use the following method. The signal at each harmonic wavelength from the experimental spectra is acquired. It is known that the harmonic signal beyond the plateau region shows as exponential decay, which results in a linear decline in the logarithmic scale. Fig.~\ref{fig:S6} depicts this clearly, where the plateau region has a constant harmonic yield followed by a linear decay in the harmonic signal. The green line indicates the average signal of the plateau harmonics and the orange line indicates a linear fit to the intensities in the cut-off region. The intersection of the green and orange lines denotes the cut-off wavelength, which is converted to $E_c$, the cut-off energy of the plateau.

\section{Theoretical details}
\subsection{\textbf{\textit{Ab initio}} supercell calculations}
\subsubsection{Molecular dynamics simulations}
In order to study theoretically the interaction of strong-field pulse with  liquid water, the first step is simulating the structural and dynamical properties of liquid water. In this regard, we use Car-Parrinello molecular dynamics (CPMD) simulations in the canonical ensemble at the temperature of 300~K. The simulated system, as displayed in the main text, is a periodic cubic supercell, including 64 \ce{H2O} molecules in the experimental liquid water density of 1~gr/cm$^2$.
Our computed radial distribution function between oxygen atoms (shown in Ref.~\cite{Zahra2021}), which is averaged over a time length of approximately 20~ps, is in a good consistent with the experimental neutron scattering and x-ray diffraction measurements; it confirms the accuracy of our liquid water molecular dynamics simulation. 
Our CPMD simulations are performed using Quantum Espresso package \cite{Giannozzi2017}. We use norm-conserving GGA-revPBE \cite{revpbe} pseudopotentials for oxygen and hydrogen atoms and Grimme-D2 dispersion correction to describe the non-local correlation effects. For more details please see Ref.~\cite{Zahra2021}. 

\subsubsection{Ultrafast dynamics simulations}
\begin{figure}[h]
  \begin{center}
    \includegraphics[width=0.95\columnwidth]{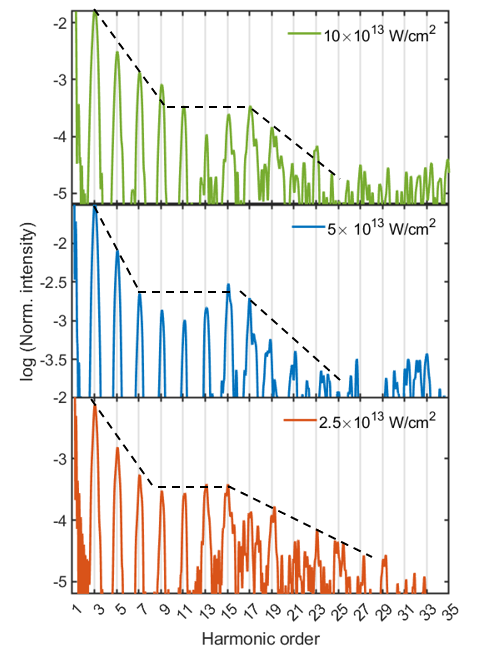}
 \caption{High-harmonic spectra obtained from calculations with the supercell approach using a driving wavelength of 1500~nm.}
 \label{fig:supercell}
 \end{center}
\end{figure}
In the next step, to study the interaction of ultrashort intense pulse with liquid target, we use real-time time-dependent density functional theory implemented in Octopus package \cite{Marques2003, Castro2003, Andrade2015}. Note that the ultrashort laser pulse stimulates the system in the time-scale of femtosecond while our CPMD simulation reproduce the macroscopic characteristics of liquid water in the time-scale of picosecond. So in order to obtain the converged response, we need to consider a sufficient number of different configurations of the liquid supercell system, denoted as 'subsystems'. The time interval between consecutive subsystems are in the order of picosecond, and the number of required subsystems for convergence depend on the target system as well as pulse characteristics (wavelength and intensity). 
In the supercell simulations, we consider the dynamics of nuclear based on Ehrenfest-TDDFT approach \cite{ehrenfest} during the interaction of ultrashort intense pulses with the liquid system. However, the effect of ionic current is negligible and the calculated HHG arises totally from electron dynamics. 
The exchange-correlation term in TDDFT calculations is described by the same GGA-revPBE pseudopotentials as those used in CPMD calculations. But, vdW interactions are not included, as they lead to no difference in the spectra.
We consider a grid spacing of 0.3 bohr thorough TDDFT calculations, and a dense k-point grid of $5\times5\times5$. Also, we use aetrs algorithm to approximate the evolution operator and the time step of 0.2~a.u. within our TDDFT calculations. 
\subsubsection{Pulse characteristics}
In order to evaluate the impact of pulse wavelength and its intensity on HHG spectra, we consider the pulse characteristics corresponding to our experimental conditions. To see the impact of driving wavelength on HHG cutoff energy, we consider two different wavelengths of 800 and 1500 nm under the constant intensity of about 20~TW/cm$^2$ (shown in Fig. 3B of the main text). Additionally, to check the intensity scaling of HHG cutoff, the pulse intensity enhances from 25 to 100~TW/cm$^2$, the pulse wavelength is constant and equal to 1500~nm (shown in Fig.~\ref{fig:supercell}).  
In all cases, the pulse duration at full width at half-maximum (FWHM) is equal to 18 fs with a sine-squared envelope shape for the vector potential.

\subsection{Cluster approach ab-initio calculations}
\begin{figure}[h]
  \begin{center}
    \includegraphics[width=0.95\columnwidth]{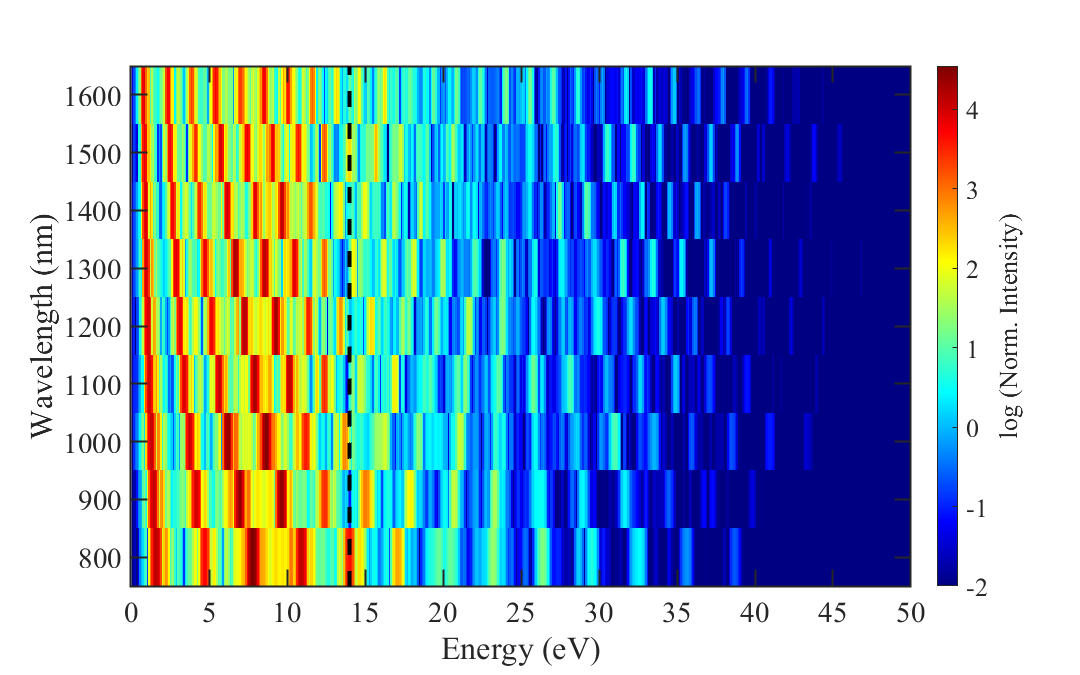}
    \includegraphics[width=0.95\columnwidth]{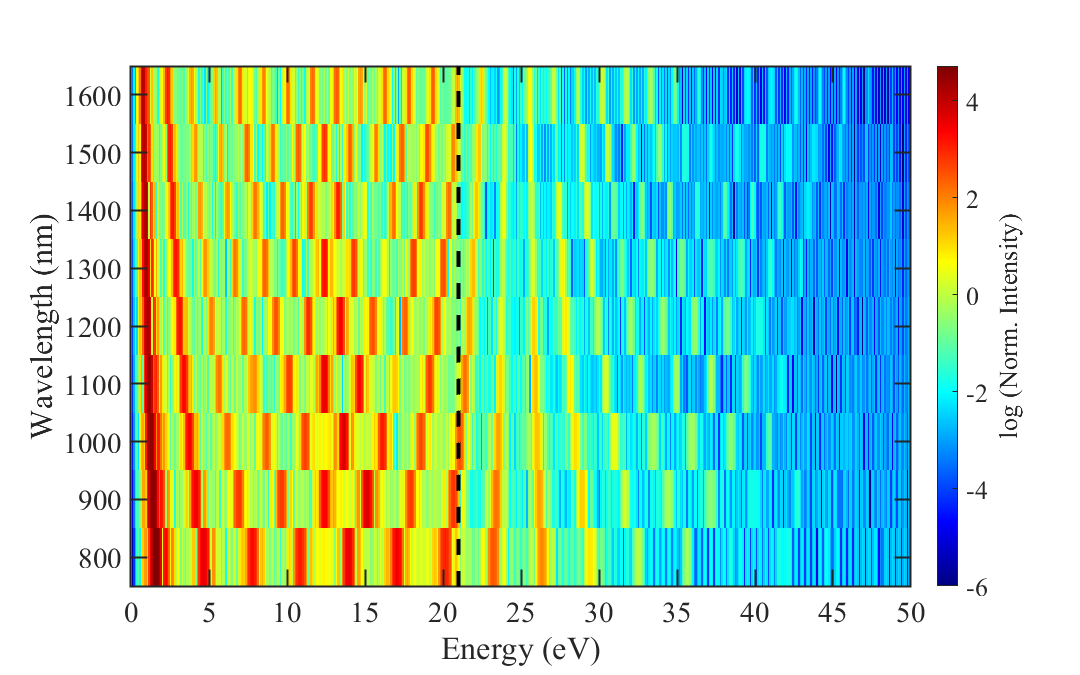}
 \caption{High-harmonic spectra obtained from calculations on ammonia (NH$_3$) clusters (top) and methane (CH$_4$) clusters (bottom) over a broad range of driving wavelengths with a peak intensity of 50~TW/cm$^2$.}
 \label{fig:clusters_CH4_NH3}
 \end{center}
\end{figure}
The numerical approach based on clusters that is utilized in the main text is thoroughly described in ref. \cite{Ofer2021}. We provide here some additional technical details for the calculations presented in the main text. All calculations were performed using the real-space grid-based code, OCTOPUS \cite{Marques2003, Castro2003, Andrade2015}. Cluster geometries for liquid \ce{H2O} were taken from ref.~\cite{Kazachenko2013} using 43-molecule large clusters for the cut-off scaling calculations, and larger 54-molecule clusters for the calculations with varying liquid density that are more sensitive. Cluster geometries for \ce{NH3} and \ce{CH4} were taken from refs.~\cite{Orabi2013} and \cite{Takeuchi2012}, respectively, with 48-molecule and 40-molecule sizes, respectively.
For each cluster, the ground state was obtained by DFT within the PBE \cite{Perdew1996} approximation for the XC functional with an added van-der-Waals correction term \cite{Grimme2010}, with a real-space cartesian grid of spacing of 0.4 Bohr and spherical grid boundaries with a radius that extended 15 Bohr beyond the farthest atom from the cluster center. In the next step, the HHG response was calculated by propagating the KS states following the prescription in ref. \cite{Ofer2021} (by suppressing contributions from the surface states, employing orientation averaging, and utilizing the independent particle approximation (the XC potential is frozen to its ground-state form)). The laser pulse was taken to have a trapezoidal envelope in all calculations with a two-cycle long turn-on and turn-off, and a four-cycle-long flat top. The dipole response was filtered with a super-Gaussian prior to obtaining the HHG spectra.

Figure \ref{fig:clusters_CH4_NH3} show high-harmonic spectra obtained with this approach for NH$_3$ clusters (top) and CH$_4$ clusters (bottom). Both panels show a cut-off energy $E_c$ that is clearly independent of the wavelength. Ammonia was chosen as an example of a different polar molecule that also forms hydrogen bonds (like water and alochols, studied in the main text) and methane was chosen as an example of an apoloar molecule. The results of Fig.~\ref{fig:clusters_CH4_NH3} thus show that the wavelength independence of $E_c$ is not specific to the liquids studied in the main text, but appears to be a general property of liquid-phase HHG.

\begin{figure}[h]
  \begin{center}
    \includegraphics[width=0.95\columnwidth]{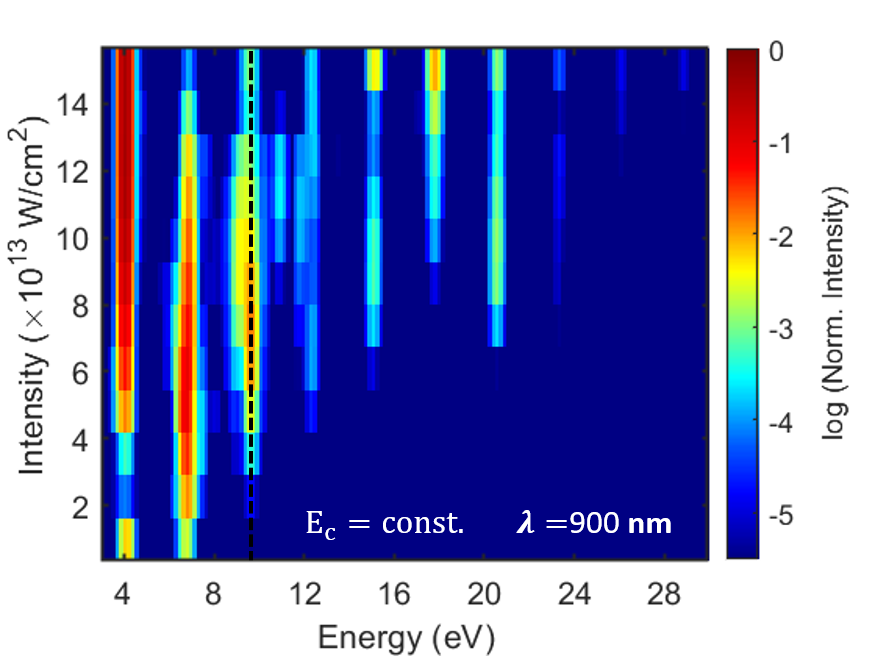}
 \caption{High-harmonic spectra obtained from calculations on water clusters using a driving wavelength of 900~nm.}
 \label{fig:clusters}
 \end{center}
\end{figure}
To complement the data shown in the main text (Figs. 3C, 3D, 4C), we show here in Fig.~\ref{fig:clusters} the scaling of the high-harmonic spectra with the peak intensity of the driving field with a central wavelength of 900~nm. The figure nicely shows the saturation of the cut-off beyond a threshold intensity.

\subsection{Extended semi-classical calculations}
\begin{figure}[h]
  \begin{center}
    \includegraphics[width=0.95\columnwidth]{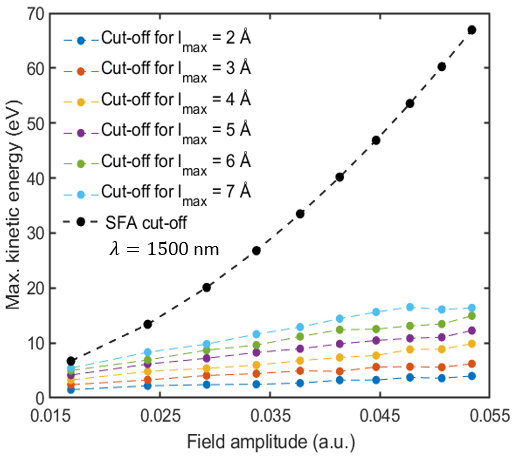}
 \caption{Scaling of the maximal electron kinetic energy with the peak electric field of the driving pulse.}
 \label{fig:tsm}
 \end{center}
\end{figure}
Based on our suggested extended semi-classical model (main text, Section II.A), we performed semi-analytical calculations for the HHG cutoff vs. the driving laser wavelength and intensity. We also performed calculations with varying $l_{max}$ parameters. In all cases, the Newton equations of motion for the electrons were solved analytically for a given time of ionization, and then represented numerically on a temporal grid to scan for re-collisions. The solutions were numerically filtered for each re-colliding trajectory to test whether that particular trajectory extruded beyond $l_{max}$ or not, while the standard TSM was not filtered with this condition. In each case the cutoff was given by the most energetic re-colliding trajectory with the added band gap of the liquid. The calculations in Fig.~\ref{fig:tsm} confirm the linear scaling of the maximal kinetic energy with the peak electric-field amplitude of the driving pulse. The onset of saturation is observed at the highest field amplitudes for $l_{\rm max}=7~{\rm\AA}$.

\subsection{Semi-classical description of HHG and cutoff scaling including scattering}
\begin{figure}[h]
  \begin{center}
    \includegraphics[width=0.95\columnwidth]{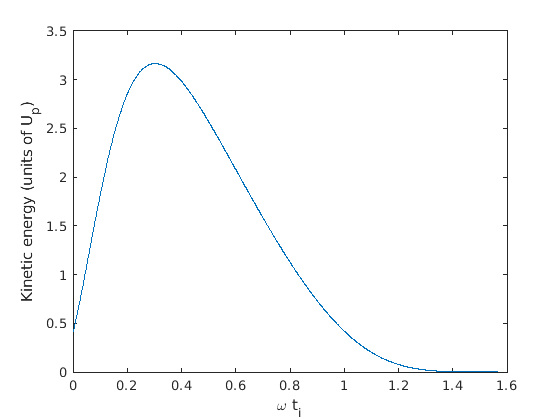}
 \caption{Evolution of the return kinetic energy versus the ionization phase, obtained from Eq.~(\ref{eq:kin}), solving numerically Eq.~(\ref{eq:tr}) for each value of $t_i$.}
 \label{fig:kin}
 \end{center}
\end{figure}

In this section, we aim at deriving an approximate fully analytical picture that describes the cutoff energy of HHG, as described by our extended semi-analytical semi-classical three-step model. We examine what is the analytical connection between the HHG cutoff and the maximum allowed excursion length of the electron trajectories in the extended model.\\
Our starting point is the full semi-classical picture discussed in the main text. We consider the classical motion of an electron in an electric field $E(t)=E\cos(\omega t)$, where $\omega$ is the frequency of the laser. The magnetic field, as well as the electric field due to the parent ion are neglected and the dipole approximation is employed.
The electron is ionized at a time $t_i$, and we consider the dynamics following this time.
\begin{eqnarray}
 x(t) &=& \frac{qE}{m\omega^2} [\cos(\omega t_i)-\cos(\omega t) - \sin(\omega t_i) \omega(t-t_i)]\,,\\
 v(t) &=& \frac{qE}{m\omega} [\sin(\omega t)-\sin(\omega t_i)] \,,
\end{eqnarray}
where we imposed as initial condition that the trajectories start at $x(t_i)=0$, with a vanishing velocity ($v(t_i)=0$).\\
The time at which the electron returns to its parent ion, referred here as the return time $t_r$, is obtained by imposing that $x(t_r)=0$.
This gives a first relation, that defines trajectories returning to the parent ion:
\begin{equation}
 [\cos(\omega t_i)-\cos(\omega t_r)] = [\sin(\omega t_i)] \omega(t_r-t_i) \,.
 \label{eq:tr}
\end{equation}
From the return time, we can obtain the kinetic energy for a trajectory starting at $t_i$
\begin{equation}
 \Delta E_K(t_i) = 2 U_p \Big[ \sin(\omega t_r) - \sin(\omega t_i) \Big]^2\,.
 \label{eq:kin}
\end{equation}
The solution of the Eq.~(\ref{eq:tr}) can be performed numerically, see Fig.~\ref{fig:kin}, from which we obtain the well known gas-phase result for the maximum kinetic energy of $\approx3.17U_p$, corresponding to $\omega t_i\approx0.305$.

Let us now consider the case that not all trajectories are allowed, as formulated in our extended model for the liquid. We first need to determine for a given trajectory, determined uniquely by $t_i$ and Eq.~(\ref{eq:tr}), what is the maximum excursion. \\
The maximum excursion is reached when $\partial_t x(t) = v(t) = 0$. This gives us the condition defining the time of maximal excursion $t^*$:
\begin{equation}
 \frac{qE}{m\omega} [\sin(\omega t^*)-\sin(\omega t_i)] = 0
\end{equation}
which has a simple analytical solution $\omega t^*=\pi-\omega t_i$.\\
We therefore have an exact expression for the excursion distance $l_{exc}$ for any given ionization time
\begin{equation}
l_{exc}(t_i) = |x(t^*)| = \frac{eE}{m\omega^2} [2\cos(\omega t_i)-\sin(\omega t_i) (\pi-2\omega t_i)]\,,
\end{equation}
with $e=|q|$ is the elementary charge.
Solving this equation numerically for the most energetic trajectory ($\omega t_i=0.306$), we find that without limiting the trajectories, the cutoff energy corresponds to an excursion of $\approx 1.145 \frac{eE}{m\omega^2}$, which verifies the analytic analysis.\\

We next consider that the maximal excursion is limited to $l_{\mathrm{max}}$.
We can distinguish two cases. As long as $1.145 \frac{eE}{m\omega^2}$ is smaller than $l_{\mathrm{max}}$, the kinetic energy gained is given by the usual formula $\Delta E_K \approx 3.17 U_p$. Above this threshold, the value of the kinetic energy gain will be modified, as the most energetic trajectory is forbidden, and the cutoff will be reduced. Below, we assume that we are in this regime.\\
The task is therefore to find the most energetic trajectory that returns to the parent ion, and whose excursion is smaller or equal to  $l_{\mathrm{max}}$.\\
This gives us the condition
\begin{equation}
 l_{exc} = \frac{eE}{m\omega^2} [2\cos(\omega t_i)-\sin(\omega t_i) (\pi-2\omega t_i)] \le l_{\mathrm{max}}\,.
 \label{cond}
\end{equation}
The function $l_{exc}(t_i)$ is a function that decreases monotonously from 2 to 0 over the interval of interest of the ionization time $[0:\pi/2]$. Therefore a value $t_i^*$ exists such that the condition (\ref{cond}) is satisfied, for all $t_i>t_i^*$ (and smaller than $\pi/2$).
Because we are in a regime where not all the trajectories are allowed, this time corresponds to a ``short trajectory'', meaning that the ionization phase is larger than $0.306$, which is the time corresponding to the most energetic trajectory without scattering. Said differently, we only select the short trajectories that have an excursion smaller than $l_{\mathrm{max}}$.
Now, we also know that the return kinetic energy decreases monotonously for the short trajectories (see Fig.~\ref{fig:kin}). Therefore, the critical time $t_i^*$ corresponds to the highest kinetic energy in our scattering model.\\
This ionization time that corresponds to the most energetic trajectory with an excursion smaller or equal to $l_{\mathrm{max}}$ is given by the relation
\begin{equation}
  [2\cos(\omega t_i^*)-\sin(\omega t_i^*) (\pi-2\omega t_i^*)] =  L_{\mathrm{red}}\,,
  \label{eq:ti}
\end{equation}
where we define the dimensionless quantity $L_{\mathrm{red}} = \frac{m\omega^2 l_{\mathrm{max}}}{eE}$.\\
We therefore have two defining equations, Eq.~(\ref{eq:ti}) and  Eq.~(\ref{eq:tr}), where the first one determines the ionization time as a function of the reduced maximum excursion $L_{\mathrm{red}}$, and the second one that determines the corresponding return time.

\begin{figure}[t]
 \begin{center}
    \includegraphics[width=0.95\columnwidth]{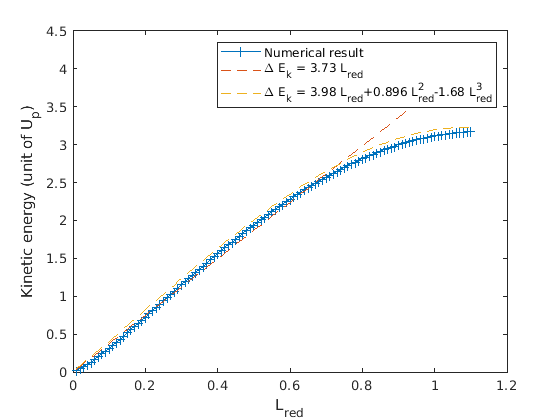}
 \caption{Evolution of the return kinetic energy versus $L_{\mathrm{red}}$. The linear and cubic fits are also shown. }
 \label{fig:fit_ek_lred}
 \end{center}
\end{figure}

We have solved numerically these two equations: for each value of $L_{\mathrm{red}}$, we can find the numerical solution of Eq.~(\ref{eq:ti}) and then use this value to numerically solve Eq.~(\ref{eq:tr}).
The result of the kinetic energy (in units of $U_p$) as a function of $L_{\mathrm{red}}$ is shown in Fig.~\ref{fig:fit_ek_lred}. For not too large values of $L_{\mathrm{red}}$, this is well approximated by a linear function.
In this regime, which is the one relevant for our experiment, we find that
\begin{equation}
 \Delta E_K(\omega t_i) \approx 3.73 U_p L_{\mathrm{red}} = \frac{3.73}{4} e E l_{\mathrm{max}}
\end{equation}

\begin{figure}[h!]
  \begin{center}
    \includegraphics[width=0.95\columnwidth]{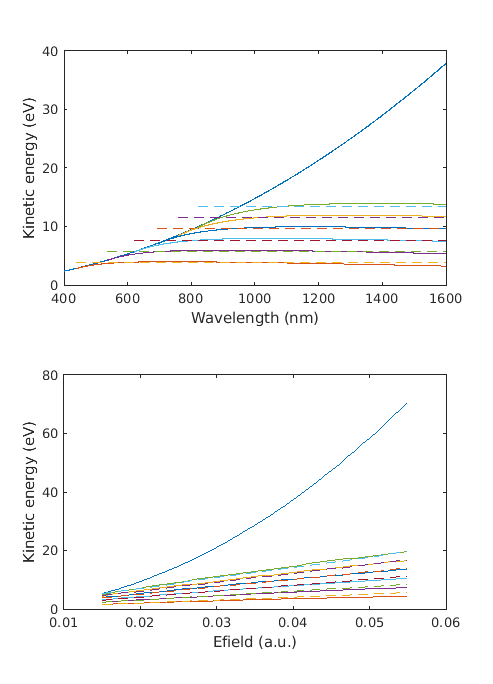}
 \caption{Evolution of the return kinetic energy versus electric field (left panel) and wavelength(right panel). The parabola correspond to the result without scattering, whereas the solid lines correspond to the result including scattering, for values of $l_{\mathrm{max}}\in\{2,4,6,8,10,12, 14\}$ Bohr. The dashed lines correspond to the same results, but obtained using the simpler formula $\Delta E_K=3.73 U_p L_{\mathrm{red}}$.}
 \label{fig:comp_model_fit}
 \end{center}
\end{figure}

From this result, we clearly see the role of introducing a mean-free path in our model. The cutoff becomes independent of the wavelength and is linear with the electric field.
The comparison between the full numerical result and the linear fit is shown for different laser parameters in Fig.~\ref{fig:comp_model_fit}, showing that for our range of parameters, the approximate linear scaling law matches very well the exact numerical result of our model. 

\section{Intensity dependence of liquid-HHG cutoff}
In the main text we concluded that the liquid-HHG cutoff was independent of the driving wavelength. However, since the HHG measurements for different wavelengths slightly vary in the driving peak laser intensity (because the threshold intensity of plasma generation depends on the laser wavelength), this conclusion is only valid if the liquid-HHG cutoff also weakly depends on the driver peak intensity in these conditions. We here show that this is indeed the case with both measurements and calculations, which validate that the HHG spectra from liquids is independent of the driving wavelength. 

\section{Comparative study of cut-off energy with different trajectory-truncation methods}

\begin{figure*} [!htbp]
\begin{centering}
\includegraphics[width=\columnwidth]{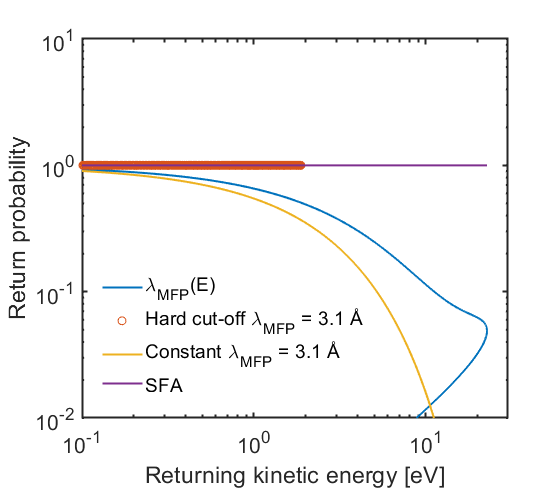}

\caption{Shows a comparison of the returning probability of trajectories with different kinetic energies for different variations of the trajectory based model. The purple solid line denotes the gas-phase classical SFA model where all trajectories upto the cut-off return with equal probability. The red circles indicate the trajectory model where returning trajectories with excursion length greater than 3.1 {\AA} are rejected. The maximum kinetic energy is therefore reduced to ~2 eV as opposed to 25 eV(SFA). The yellow line indicates further sophistication of the trajectory model where each returning trajectory is modulated by a scattering amplitude of exp(-$l_{exc}$/3.1{\AA}). This model considers exponential damping  by a fixed $\lambda_{MFP}$ value of 3.1 {\AA}. Finally the blue solid line indicates the returning probability for trajectories calculated considering energy dependent MFP. In this model at each time step of each trajectory, based on their instantaneous kinetic energy,  a scattering amplitude of exp(-$l_{exc}$/$\lambda_{MFP}(E)$) is applied to the returning probability.
}
\label{fig:S10}
\end{centering}
\end{figure*}

To study the dependence of the harmonic energy on the implementation of the electron mean free paths, we compare the cut-off energies obtained from different truncation methods.  Fig.~\ref{fig:S10} shows the variation of harmonic yield (i.e. the number of returning trajectories) as a function of the kinetic energy. The analysis is run for different models to find the harmonic yield dependence on the harmonic energy due to the energy dependent $\lambda_{\rm MFP}$. The simulations are performed with a 1500 nm light of 1.6V/{\AA} electric field amplitude. The purple solid line denotes the gas-phase classical SFA model where all trajectories upto the cut-off return with equal probability. The red circles indicate the trajectory model where returning trajectories with excursion lengths greater than 3.1 {\AA} are rejected. The maximum kinetic energy is therefore reduced to $\sim$2 eV as opposed to 25 eV (SFA).  However, in reality trajectories with trajectory lengths extending beyond the $\lambda_{\rm MFP}$ are expected to return but with exponentially decaying probability.
Indeed, the probability that a particle is scattered between the travelled distance $l$ and $l+dl$ is given by
\begin{equation}
 d\mathcal{P}(l) = \frac{1}{\lambda_{\rm MFP}} e^{-l/\lambda_{\rm MFP}} dl\,,
\end{equation}
where $\lambda_{\rm MFP}$ is the mean free path of the particle.
From this definition, we can compute the probably that a particle is getting scattered after a travelled distance $L$, given by the travelled distance at the returned time. This is given by
\begin{equation}
 \mathcal{P}(L) = \int_{0}^L \frac{1}{\lambda_{\rm MFP}} e^{-l/\lambda_{\rm MFP}} dl = (1 - e^{-L/\lambda_{\rm MFP}})\,.
\end{equation}
As $L$ grows, the probability converges to one, as one would naturally expect.
In this treatment, we have assumed that the mean free path is independent of the energy of the particle, and therefore the scattering probably only depends on the travelled distance. \\
It is however possible to go beyond this and to take into account the whole variation of the mean free path along the path of the electron, as we know analytically the trajectories.
For this, we convert the integral over the travelled distance into an integral over the time, using the fact that the travelled distance at a point in time $t$ is given by 
$l(t) = \int_{t_i}^t |v(t')| dt'$.
\begin{equation}
 \mathcal{P}(L) = \int_{t_i}^{t_r} \frac{1}{\lambda_{\rm MFP}(E_K(t))} e^{-l(t)/\lambda_{\rm MFP}(E_K(t))} |v(t)| dt \,.
\end{equation}

 The yellow line indicates further sophistication of the trajectory model as given by equation (11). This model considers exponential damping  by a fixed $\lambda_{MFP}$ value of 3.1 {\AA}. Finally the blue solid line indicates the returning probability for trajectories calculated considering energy-dependent MFP given by equation (12). We observe that even with the statistical Monte-Carlo-like energy-dependent MFP model (solid blue line), the harmonic yield (returning probability) varies slowly up to $\sim$ 2 eV falling to only 0.5 the value with respect to the hard-cut off and the SFA model. Beyond $\sim$ 2 eV the harmonic yield falls sharply indicating the onset of the cut-off region. This is in considerable agreement with the simplest hard-cut-off model. As a result, the harmonic yield below cut-off does not change considerably to extract the energy-dependent  $\lambda_{MFP}$(E) with reasonable accuracy taking into account the effect of experimental parameters. Beyond the cut-off energy there is a significant dependence of the harmonic yield on the harmonic energy resulting from $\lambda_{MFP}$(E). However, the above cut-off harmonics also have contributions from back scattering and nearest-neighbor recombination as discussed previously. Thus isolating the effect of the energy-dependent $\lambda_{MFP}$ only for above-cut-off harmonics is beyond the scope of this manuscript. 

\section{Analytical treatment of the propagation effect}

Our goal is to describe how the laser-induced current density lead light emission at a given frequency propagate in a medium, cross the surface of the material and reach the detector. 
We first calculate the $\mathbf{E}$ and $\mathbf{B}$ fields that result from given source terms. 
Our approach is quite similar to the approach of Ref.~\onlinecite{Sipe_Green_functions}, that we generalize here to not only treat the polarization, but also the free charge current and densities, denoted respectively $\rho(\mathbf{r},t)$ and  $\mathbf{J}$. 
In the context of high-harmonic generation in solids, it is mandatory to keep these terms, as the free-charge current ($\mathbf{J}$) describes the intraband contribution while the polarization $\mathbf{P}$, describes the interband contribution.

We begin with the macroscopic Maxwell equations for an homogeneous system
\begin{eqnarray}
 \boldsymbol{\nabla}\mathbf{D}(\mathbf{r},t) &=& 4\pi\rho(\mathbf{r},t) \nonumber\\
 c\boldsymbol{\nabla}\times\mathbf{H}(\mathbf{r},t)-\frac{\partial \mathbf{D}}{\partial t}(\mathbf{r},t)&=&4\pi\mathbf{J}(\mathbf{r},t), \nonumber\\
  \boldsymbol{\nabla}\mathbf{B}(\mathbf{r},t) &=& 0, \nonumber\\
  c\boldsymbol{\nabla}\times\mathbf{E}(\mathbf{r},t)+ \frac{\partial \mathbf{B}}{\partial t}(\mathbf{r},t) &=&0,
\end{eqnarray}
where 
\begin{eqnarray}
 \label{macroconst}
 \mathbf{D}(\mathbf{r},t) = \mathbf{E}(\mathbf{r},t) + 4\pi\mathbf{P}_{\mathrm{tot}}(\mathbf{r},t),\nonumber\\
 \mathbf{H}(\mathbf{r},t) = \mathbf{B}(\mathbf{r},t) - 4\pi\mathbf{M}_{\mathrm{tot}}(\mathbf{r},t).
\end{eqnarray}

The total macroscopic polarization, $\mathbf{P}_{\mathrm{tot}}$, and magnetization, $\mathbf{M}_{\mathrm{tot}}$, have in frequency space the following expression
\begin{eqnarray}
 \mathbf{P}_{\mathrm{tot}}(\mathbf{r},\omega) = \chi_e(\omega) \mathbf{E}(\mathbf{r},\omega) + \mathbf{P}_{\mathrm{nl}}(\mathbf{r},\omega) \nonumber\\
 \mathbf{M}_{\mathrm{tot}}(\mathbf{r},\omega) = \chi_b(\omega) \mathbf{B}(\mathbf{r},\omega) + \mathbf{M}_{\mathrm{nl}}(\mathbf{r},\omega),
\end{eqnarray}
where the macroscopic susceptibilities $\chi_e$ and $\chi_b$ describe the linear response of the medium, and are directly related to the macroscopic dielectric and the permeability tensors of the system
\begin{eqnarray}
 \epsilon(\omega) = 1+4\pi \chi_e(\omega), \nonumber\\
 \mu^{-1}(\omega) = 1-4\pi \chi_b(\omega). \nonumber
\end{eqnarray}
Note that as we are interested here in systems which are homogeneous at equilibrium, these quantities do not depend on space.
The remaining polarization $\mathbf{P}_{\mathrm{nl}}(\mathbf{r},\omega)$ (magnetization $\mathbf{M}_{\mathrm{nl}}(\mathbf{r},\omega)$) term corresponds to the nonlinear contributions to the total polarization (magnetization).

Rewriting the Maxwell equations only in terms of $\mathbf{E}$ and $\mathbf{B}$ fields, we obtain
\begin{eqnarray}
 \epsilon(\omega)\boldsymbol{\nabla}\mathbf{E}(\mathbf{r},\omega) &=& 4\pi n_{\mathrm{tot}}(\mathbf{r},\omega)   \nonumber\\
 c\boldsymbol{\nabla}\times\mathbf{B}(\mathbf{r},\omega) -i\omega\epsilon(\omega)\mu(\omega)\mathbf{E}(\mathbf{r},\omega) &=&4\pi\mu(\omega)\mathbf{j}_{\mathrm{tot}}(\mathbf{r},\omega)  ,\nonumber\\
  \boldsymbol{\nabla}\mathbf{B}(\mathbf{r},\omega) &=& 0, \nonumber\\
  c\boldsymbol{\nabla}\times\mathbf{E}(\mathbf{r},\omega)+ i\omega\mathbf{B}(\mathbf{r},\omega) &=&0,
  \label{eqs:Maxwell_E_B}
\end{eqnarray}
where we defined the total macroscopic nonlinear current and charge densities as
\begin{eqnarray}
 n_{\mathrm{nl}}(\mathbf{r},\omega) = \rho(\mathbf{r},\omega) -\boldsymbol{\nabla}\mathbf{P}_{\mathrm{nl}}(\mathbf{r},\omega), \nonumber\\
 \mathbf{j}_{\mathrm{nl}}(\mathbf{r},\omega) = \mathbf{J}(\mathbf{r},\omega)  + i\omega\mathbf{P}_{\mathrm{nl}}(\mathbf{r},\omega)+ c \boldsymbol{\nabla}\times\mathbf{M}_{\mathrm{nl}}(\mathbf{r},\omega).
\end{eqnarray}
These are the source terms induced by the strong incoming laser, which in term induce the harmonic fields.\\
There are many situations where the spatial variations of the current cannot be neglected. 
Due to the multi-photon absorption, the pump laser is depleted, which implies a spatial variation of the laser intensity, and hence of the generated current along the propagation direction. 
As the laser-matter interaction is highly nonlinear in the case of high-order harmonic generation, many effects have to be taken into account when describing the propagation of the pump laser in the matter, such as self-focusing effects~\cite{boyd2003nonlinear}, which is caused by the third-order non-linearity of the medium.
Moreover, due to surface roughness, nano-structuration of the surface, or possible metalization of the first few atomic layers at the incident surface, the incoming laser can excite surface currents and surface-plasmon-polaritons when the field crosses the interface~\cite{2040-8986-18-11-115007}.
It is clear that all these phenomena will induce a spatial variation of the pump laser will propagating in the matter, even at normal incidence, and will thus induce a spatial dependence of the generated nonlinear current. However, while our treatment is general and does no require any approximation of the spatial dependence of the current, we will in the following assume that we have a film which is thin compared to the Rayleigh length of the driving pulse, and neglect the spatial dependence and propagation effects of the incoming laser field.\\
We now proceed, following the approach of Ref.~\onlinecite{Sipe_Green_functions} and we first give the solution of the homogeneous problem defined by Eqs.~\ref{eqs:Maxwell_E_B}, without considering the presence of the surface.
In the view of treating a surface, we construct plane-wave solutions of Eqs.~\ref{eqs:Maxwell_E_B} which have a real momentum in the ($xy$) plane of the surface.~\cite{Sipe_Green_functions}
The $z$-axis is chosen to be perpendicular to the surface plane.\\

\begin{figure}
 \begin{centering}
 \includegraphics[width=0.7\columnwidth]{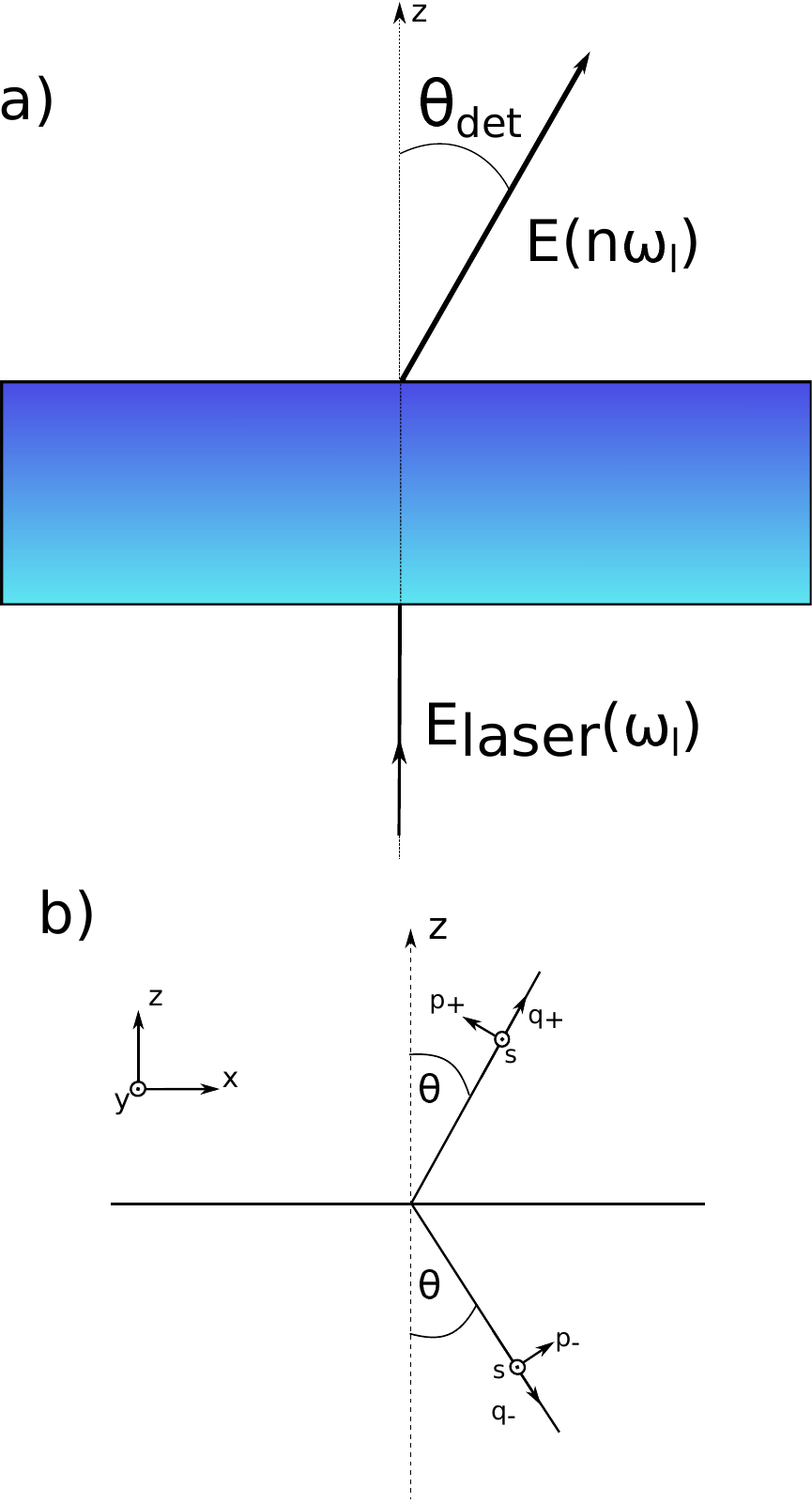}
 \caption{\label{fig:sketch} (a)The experimental geometry considered in this work. The incoming laser field $\mathbf{E}_{\mathrm{laser}}$ is at normal incidence. (b) Coordinate systems used to describe propagation of harmonics. The unit vectors and the link between the different coordinate systems are given in the text. The sketch is presented here for an azimuthal angle $\phi=0$. Adapted from Ref.~\onlinecite{Sipe_Rpp}}
 \end{centering}
\end{figure}
There is therefore two possible wave-vectors associated
 \begin{eqnarray}
      \mathbf{q}_+(\omega) &=& q_{||}(\omega)\boldsymbol{\hat{\kappa}} + q_\bot(\omega)\mathbf{\hat{z}}\,, \nonumber\\
      \mathbf{q}_-(\omega) &=& q_{||}(\omega)\boldsymbol{\hat{\kappa}} - q_\bot(\omega)\mathbf{\hat{z}}\,,\nonumber
 \label{eq:wavevectors}
\end{eqnarray}
where $\boldsymbol{\hat{\kappa}}$ is the in-plane unit vector of the form $\boldsymbol{\hat{\kappa}} = \cos(\phi)\mathbf{\hat{x}} + \sin(\phi)\mathbf{\hat{y}}$, $\phi$ being the azimuthal angle, and $q_\bot(\omega)$ is the out-of-plane part of the wave-vector. \\
Here $q(\omega)^2 = \mu(\omega)\epsilon(\omega)\left(\frac{\omega}{c}\right)^2 = n(\omega)^2\left(\frac{\omega}{c}\right)^2$.
and $q_\bot(\omega)=\left(\mu(\omega)\epsilon(\omega)\frac{\omega^2}{c^2}-q_{||}^2(\omega)\right)^{1/2}=\frac{\omega}{c}\left(\mu(\omega)\epsilon(\omega)-\sin^2(\theta)\right)^{1/2}$, $q_\bot$ being chosen to have $\mathrm{Im}[q_\bot]\geq0$ and $\mathrm{Re}[q_\bot]\geq0$ if $\mathrm{Im}[q_\bot]=0$.
This convention allows us to be sure that the wave associated with $\mathbf{q}_+$ is upward propagating and that the one associated with $\mathbf{q}_-$ is downward-propagating.~\cite{Sipe_Rpp,Sipe_Green_functions} 
We thus refer to the solutions of the homogeneous problem associated with $\mathbf{q}_+$ (respec. $\mathbf{q}_-$) as the  upward-propagating (respec. downward-propagating) solutions.\\
The treatment of the problem represented by equation Eqs.~\ref{eqs:Maxwell_E_B} if greatly simplified if we introduce the $s$ and $p$ polarizations, defined by the following unit vectors
\begin{eqnarray}
 \mathbf{\hat{s}} &=& \boldsymbol{\hat{\kappa}}\times\mathbf{\hat{z}}\,,\nonumber\\
 \mathbf{\hat{p}}_{\pm}(\omega) &=&  \frac{(q_{||}(\omega)\mathbf{\hat{z}} \mp q_\bot(\omega) \boldsymbol{\hat{\kappa}})}{q(\omega)}.\nonumber
\end{eqnarray}
This is illustrated in Fig.~\ref{fig:sketch}(b). For conciseness, we have omitted the frequency dependence here.\\
One can easily check that we have the following relations
 \begin{equation}
\begin{aligned}
     \mathbf{\hat{s}}\times\mathbf{\hat{q}}_{\pm}(\omega) =& \mathbf{\hat{p}}_{\pm}(\omega)\,, \nonumber \\
      \mathbf{\hat{q}}_{\pm}(\omega)\times\mathbf{\hat{p}}_{\pm}(\omega) =& \mathbf{\hat{s}}\,, \nonumber  \\
     \mathbf{\hat{p}}_{\pm}(\omega)\times\mathbf{\hat{s}} =& \mathbf{\hat{q}}_{\pm}(\omega)\,, \nonumber 
\end{aligned}
\end{equation}
where $\mathbf{\hat{q}}_{\pm}(\omega)=\mathbf{q}_{\pm}(\omega)/q(\omega)$. Thus $(\mathbf{\hat{s}};\mathbf{\hat{q}}_+;\mathbf{\hat{p}}_+)$ and $(\mathbf{\hat{s}};\mathbf{\hat{q}}_-;\mathbf{\hat{p}}_-)$ are direct basis.\\
Therefore, the wave associated with  $\mathbf{\hat{q}}_{+}$ (respectively $\mathbf{\hat{q}}_{-}$) only propagates along $\mathbf{\hat{s}}$ and $\mathbf{\hat{p}}_{+}$ (respectively $\mathbf{\hat{p}}_{-}$).\\
Therefore, the upward-propagating solutions of the homogeneous Maxwell equations is of the form
 \begin{equation}
\begin{aligned}
 \mathbf{E}_{+}(\mathbf{r},\omega) =& \left[E_{s^+}(\omega)\mathbf{\hat{s}}+E_{p^+}(\omega)\mathbf{\hat{p}_+}(\omega)\right]e^{i\mathbf{q_+}(\omega)\ldotp\mathbf{r}}\,,\\
 \mathbf{B}_{+}(\mathbf{r},\omega) =& n(\omega)\left[E_{p^+}(\omega)\mathbf{\hat{s}}+E_{s^+}(\omega)\mathbf{\hat{p}_+(\omega)}\right]e^{i\mathbf{q_+}(\omega)\ldotp\mathbf{r}}\,.\\
\end{aligned}
\label{eq:E_B_plus_fields}
\end{equation}
The downward-propagating solution as the same form and is not reported here for conciseness.\\
Knowing the solution of the homogeneous problem, we can now search for the solution of the full problem, described by Eqs.~\ref{eqs:Maxwell_E_B}.
The search of the solution of Eqs.~\ref{eqs:Maxwell_E_B} proceed in two steps. We first find the solution associated with source terms of the form~\cite{Sipe_Green_functions}
 \begin{equation}
 \mathbf{j}_{\mathrm{tot}}(\mathbf{r},\omega)  = \mathbf{j}_{\mathrm{tot}}(q_{||},z_0,\omega)\delta(z-z_0)e^{iq_{||}\ldotp\mathbf{r}_{||}}\,,
\label{eq:source_terms_sheet}
\end{equation}
with $\mathbf{r}_{||} = (x,y)$. This corresponds to a sheet of current and density located in the plane $z=z_0$ and having a in-plane wave-vector $q_{||}(\omega)$.
The physical solution of the Maxwell equations, for such a source term located at $z=z_0$ (which excludes exponentially diverging waves)  still fulfill the homogeneous Maxwell equations for $z\neq z_0$ 
and therefore reads~\cite{Sipe_Green_functions}
\begin{eqnarray}
     \mathbf{E}(\mathbf{r};\omega) = \Big[\mathbf{E}_+(\mathbf{r};\omega)\Theta(z-z_0)e^{-iq_\bot(\omega) z_0}\nonumber\\
     +\mathbf{E}_-(\mathbf{r};\omega)\Theta(z_0-z)e^{iq_\bot(\omega) z_0} \nonumber\\
     + \mathcal{E}(\omega)\delta(z-z_0)e^{iq_{||}(\omega)\ldotp\mathbf{r}_{||}}\Big]\,,
\end{eqnarray}
and a similar expression for the magnetic field, with here $\Theta(z)$ being the Heaviside function.\\
The determination of the coefficients $E_{s^\pm}(\omega)$ and $E_{p^\pm}(\omega)$ of Eq.~\eqref{eq:E_B_plus_fields} proceed exactly as explained in details in Ref.~\onlinecite{Sipe_Green_functions}, and is therefore not reported here.
After some algebra, we obtain that
 \begin{equation}
\begin{aligned}
  {E}_{s\pm} =& \frac{2\pi \mu(\omega) \omega }{q^m_\bot(\omega) c^2}\mathbf{\hat{s}}\ldotp\mathbf{j}_{\mathrm{tot}}(\mathbf{q_{||}},z_0,\omega)\,,\\
  {E}_{p\pm} =& \frac{2\pi  \mu(\omega) \omega}{ q^m_\bot(\omega) c^2 } \mathbf{\hat{p}}^m_{\pm}(\omega)\ldotp\mathbf{j}_{\mathrm{tot}}(\mathbf{q_{||}},z_0,\omega)\,,\\
  \mathcal{B}(\omega) =& 0, \qquad \mathcal{E}(\omega) = -\frac{4\pi i}{\omega \epsilon(\omega)} \mathbf{\hat{z}} \mathbf{\hat{z}}\ldotp\mathbf{j}_{\mathrm{tot}}(\mathbf{q_{||}},z_0,\omega)\,.
\end{aligned}
\label{eq:phenom_fiel}
\end{equation}

These coefficients determine completely the electric and magnetic fields induced by the source terms of Eqs.~\eqref{eq:source_terms_sheet}. 

The total electric field induced by one sheet of polarization located at $z_0$, and evaluated at the position $\mathbf{r}$, with a given in-plane wave-vector $\mathbf{q_{||}}$, denoted $\mathbf{E}_{\mathbf{q_{||}},z_0}$, is thus given by
\begin{widetext}
 \begin{eqnarray}
     \mathbf{E}_{\mathbf{q_{||}},z_0}(\mathbf{r};\omega) = \Big[
     \frac{2\pi \mu(\omega) \omega }{q^m_\bot(\omega) c^2}\left[\mathbf{\hat{s}}\mathbf{\hat{s}}+\mathbf{\hat{p}}^m_{+}(\omega)\mathbf{\hat{p}}^m_{+}(\omega)\right] \Theta(z-z_0)e^{iq_\bot(\omega) (z-z_0)}\nonumber\\
     +\frac{2\pi \mu(\omega) \omega }{q^m_\bot(\omega) c^2}\left[\mathbf{\hat{s}}\mathbf{\hat{s}}+\mathbf{\hat{p}}_{-}^m(\omega)\mathbf{\hat{p}}_{-}^m(\omega)\right] \Theta(z_0-z)e^{iq_\bot(\omega) (z_0-z)} 
     - \frac{4\pi  i}{\omega \epsilon(\omega)} \mathbf{\hat{z}} \mathbf{\hat{z}}\delta(z-z_0)\Big]\mathbf{j}_{\mathrm{tot}}(\mathbf{q_{||}},z_0,\omega)e^{iq_{||}\ldotp\mathbf{r}_{||}}\,.
\end{eqnarray}
\end{widetext}
A similar expression holds for the $\mathbf{B}$ field.
The first term of this expression corresponds to the upward propagating field radiating from the considered source term sheet. The second term is the downward propagating field, and the last one is the local contribution that does not radiate~\cite{Sipe_Green_functions}. \\
Exploiting the linearity of Maxwell equations (Eqs.~\ref{eqs:Maxwell_E_B}), we can obtain the total electric field originating from any general source terms $n_{\mathrm{tot}}(\mathbf{r},\omega)$ and $\mathbf{j}_{\mathrm{tot}}(\mathbf{r},\omega)$
by integrating over the corresponding in-plane wave-vector and $z_0$, the latter ranging from $z_b$ to $z_t$. \\
The general fields $\mathbf{E}(\mathbf{r};\omega)$ are obtained by summing the individual contributions from all the source terms sheets
\begin{equation}
  \mathbf{E}(\mathbf{r};\omega) =  \int \frac{d^2 \mathbf{q_{||}}}{2\pi} \int_{z_b}^{z_t} dz_0 \mathbf{E}_{\mathbf{q_{||}},z_0}(\mathbf{r},\omega)\,.
\end{equation}
Similarly, one can also defined the total upward propagating electric field and the magnetic field. After some algebra and using Eq.~\eqref{eq:source_terms_sheet}, we obtain
\begin{eqnarray}
  \mathbf{E}_{+}(\mathbf{r};\omega) =  \frac{2\pi \mu(\omega) \omega }{q^m_\bot(\omega) c^2}\left[\mathbf{\hat{s}}\mathbf{\hat{s}}+\mathbf{\hat{p}}^m_{+}(\omega)\mathbf{\hat{p}}^m_{+}(\omega)\right]  \nonumber\\
  \times  
  \int_{z_b}^{z} dz_0 \mathbf{j}_{\mathrm{tot}}(\mathbf{r_{||}},z_0,\omega)e^{iq_\bot(\omega) (z-z_0)}.
\end{eqnarray}
Similar expressions are obtained for the downward propagating electric field and the magnetic upward/downward propagating fields.
In the following, we assume that we can neglect the in-plane position dependence of the electronic current. Thus
\begin{eqnarray}
  \mathbf{E}_{+}(z;\omega) &=&  \frac{2\pi \mu(\omega) \omega }{q^m_\bot(\omega) c^2}\left[\mathbf{\hat{s}}\mathbf{\hat{s}}+\mathbf{\hat{p}}^m_{+}(\omega)\mathbf{\hat{p}}^m_{+}(\omega)\right]  \nonumber\\
  &&\times  
  \int_{z_b}^{z} dz_0 \mathbf{j}_{\mathrm{tot}}(z_0,\omega)e^{iq_\bot(\omega) (z-z_0)}.
\end{eqnarray}

\subsection{Measured intensity for thick films}

From the previous results, we can now obtain the expression of the radiated harmonic field in the entire system.
However, we are only interested in the transmitted harmonic field, which corresponds to the to the electric field in vacuum, measured at some value of $z$ such that $z>z_t$. 
As we are interested by a macroscopic treatment, the surfaces are modeled by abrupt surfaces with effective Fresnel coefficients. 
These effective coefficients can be always chosen such that they include the microscopic details of the surfaces.
At each interface, a Maxwell saltus occur and only the in-plane component of the electric field's wave-vector is conserved.
Therefore, choosing the polarization and wave-vector of the incoming field determines uniquely the fields in vacuum.\\
We first neglect the multiple reflection, assuming a thick slab. The case of a thin-film with multiple reflections is discussed in the next section.\\
Knowing the expression of the upward-propagating harmonic fields, we can now look at the expression of the transmitted harmonic light in vacuum. \\
The harmonic electric field emitted by the crystal is considered to be weak enough to not induce new nonlinear effects, and to have its propagation well described by linear optics laws. At the interface with the vacuum, the field is therefore modified according to the Fresnel coefficients.
The harmonic electric field that reaches the detector is thus $\mathbf{E}_{\text{out}}(z;n\omega_l) = \mathbf{E}_{\text{out}}(z=z_t^+;n\omega_l)e^{iq_\bot^v(n\omega_l)(z-z_t)}=\mathbf{E}_{\text{out}}(n\omega_l)e^{iq_\bot^v(n\omega_l)(z-z_t)}$, with
\begin{equation}
 \mathbf{E}_{\text{out}}(n\omega_l) = \mathbf{\hat{e}}^{n\omega_l} \ldotp \mathbf{E_+}(z_t^-;n\omega_l)\,,
 \label{eq:transmission_simple}
\end{equation}
where $\mathbf{E_+}(z_t^-;n\omega_l)$ is the upward propagating field computing just below the surface.
%
%
Here, we have defined $\mathbf{\hat{e}}^{n\omega_l} = \mathbf{\hat{e}^{out}}(n\omega_l) \left[\mathbf{\hat{s}} t^s_{mv}(n\omega_l)\mathbf{\hat{s}} + \mathbf{\hat{p}}^v_{+}(n\omega_l) t^p_{mv}(n\omega_l)\mathbf{\hat{p}}^m_{+}(n\omega_l)\right]$, $\mathbf{\hat{e}^{out}}(n\omega_l)$ being the measured polarization. Note that the vectors $ \mathbf{\hat{p}}^i_{+}(n\omega_l) $ are defined with the detection angle $\theta_{\mathrm{det}}$ and not the incidence angle.
 
\begin{widetext}
If we assume no pump depletion and neglect nonlinear effects in the pump propagation, we can drop position dependence in source term, the electronic current, and replace it by the power $n$ of the phase of the incoming laser, which leads to

\begin{eqnarray}
 \mathbf{E}_{\text{out}}(n\omega_l) &=& \frac{2\pi \mu(n\omega_l) n\omega_l }{q^m_\bot(n\omega_l) c^2}\mathbf{\hat{e}}^{n\omega_l}\ldotp 
  \mathbf{j}_{\mathrm{tot}}(n\omega_l) e^{i(q_\bot(n\omega_l)z_t-nq_\bot(\omega_l)z_b)}
  \int_{z_b}^{z_t} dz_0  e^{i(nq_\bot(\omega_l)-q_\bot(n\omega_l))z_0} \nonumber\\
  &&=  \frac{2\pi \mu(n\omega_l) n\omega_l }{q^m_\bot(n\omega_l) c^2}\mathbf{\hat{e}}^{n\omega_l}\ldotp 
  \mathbf{j}_{\mathrm{tot}}(n\omega_l) 
  i\frac{e^{iq_\bot(n\omega_l)L}-e^{inq_\bot(\omega_l)L}}{(nq_\bot(\omega_l)-q_\bot(n\omega_l))}\,,
\end{eqnarray}
where the last term is the phase-matching term and $L=z_t-z_b$. This implies here that we are in the perturbative regime, i.e., that the $n$-th harmonic scales as the $n$-th power of the electric field. The modification of the refractive index at the incoming frequency, due to the Kerr effect for instance, can still be taken into account in $q_\bot(\omega_l)$. Note that experimental values points toward a dependence in the electric field of the $n$-th harmonic much lower that the $n$-th power of the electric field, and one might replace $nq_\bot(\omega_l) $ by $n_{\mathrm{eff}}q_\bot(\omega_l)$ where $n_{\mathrm{eff}}$ is an effective number for all harmonic order.
From this expression, the measured intensity of the harmonic field is given by the usual relation $I(n\omega_l) = \frac{c\epsilon_0}{2}| \mathbf{E}_{\text{out}}(n\omega_l)|^2$.\\
One special case of interest is normal incidence. As we assumed a cubic, non-magnetic crystal, a $s$($p$)-polarized driving field lead to a $s$($p$)-polarized current, and thus $ I_{pp}(n\omega_l) =  I_{ss}(n\omega_l) = I(n\omega_l)$, 
with
\begin{eqnarray}
 I(n\omega_l) = \frac{2\epsilon_0\pi^2}{c\epsilon(n\omega_l)}\left|\mathbf{\hat{e}}^{n\omega_l}\ldotp\mathbf{j}_{\mathrm{tot}}(n\omega_l) 
  \right|^2\left|\frac{e^{iq_\bot(n\omega_l)L}-e^{inq_\bot(\omega_l)L}}{(nq_\bot(\omega_l)-q_\bot(n\omega_l))}\right|^2\nonumber\\
  \label{eq:int_normal_incidence_simple}
\end{eqnarray}
This expression is the generalization of the typical phase-matching expression that one would obtain from perturbative non-linear optic to the $n$-th order harmonic generation from a slab of matter. It contains the propagation of the incoming light pulse from $z_b$ to $z$ and then the subsequent propagation of the emitted light from $z$ to $z_t$, for all possible values of $z$.

\end{widetext}

\subsection{Measured intensity for thin films}

In the case of a thin-film, it is not possible to neglect the multiple reflections of the harmonics inside the material. In this case, Eq.~\ref{eq:transmission_simple} has to be modified, to account for internal reflections. The resulting total outgoing field is composed of the transmitted upward propagating field, plus the multiple reflection of the upward and downward propagating waves.

\begin{widetext}
 
\begin{eqnarray}
 \mathbf{E}_{\text{out}}(n\omega_l) &=&  \frac{2\pi \mu(n\omega_l) n\omega_l }{q^m_\bot(n\omega_l) c^2}\mathbf{\hat{e}}^{n\omega_l} \ldotp  \int_{z_b}^{z_t} dz_0 \mathbf{j}_{\mathrm{tot}}(z_0,n\omega_l)e^{iq_\bot(n\omega_l) (z_t-z_0)} \nonumber\\
        &&+ \frac{2\pi \mu(n\omega_l) n\omega_l }{q_\bot(n\omega_l) c^2}\mathbf{\hat{e}}^{n\omega_l}\ldotp (\mathbf{\hat{r}_+}\ldotp\mathbf{\hat{r}_-})\ldotp\sum_{n\ge 1} \left(e^{iq_\bot^m(n\omega_l)L}\right)^{2n} (\mathbf{\hat{r}_+}\ldotp\mathbf{\hat{r}_-})^{(n-1)} 
        \ldotp \int_{z_b}^{z_t} dz_0 \mathbf{j}_{\mathrm{tot}}(z_0,n\omega_l)e^{iq_\bot(n\omega_l) (z_t-z_0)}  \nonumber\\
        && + \frac{2\pi \mu(n\omega_l) n\omega_l }{q^m_\bot(n\omega_l) c^2}\mathbf{\hat{e}}^{n\omega_l}\ldotp \mathbf{\hat{r}_+}\ldotp\sum_{n\ge 1} \left(e^{iq_\bot(n\omega_l)L}\right)^{(2n-1)}  (\mathbf{\hat{r}_-}\ldotp\mathbf{\hat{r}_+})^{(n-1)}
        \ldotp
        \int_{z_b}^{z_t} dz_0 \mathbf{j}_{\mathrm{tot}}(z_0,n\omega_l)e^{iq_\bot(n\omega_l) (z_0-z_b)} 
\end{eqnarray}
where $L=z_t-z_b$ is the thickness of the thin-film and $\mathbf{\hat{r}_\pm} = [\mathbf{\hat{s}}r^s_{mv}(n\omega_l)\mathbf{\hat{s}} + \mathbf{\hat{p}}^m_{\pm}(n\omega_l)r^p_{mv}(n\omega_l)\mathbf{\hat{p}}^m_{\mp}(n\omega_l)]$.
 Following the same approach as in previous section, we now neglect the position dependence of the current and assume a perturbative dependence on the propagation phase of the incoming laser pulse.
In the case of normal incidence, assuming again a cubic non-magnetic material, the $s$ and $p$ polarization give the same result and the intensity of the radiated field is then given by
\begin{eqnarray}
 I(n\omega_l) = \frac{2\epsilon_0\pi^2}{c\epsilon(n\omega_l)}\left|\mathbf{j}_{\mathrm{tot}}(n\omega_l) 
  \right|^2 \Bigg|  
        \frac{t_{mv}}{1-e^{2iq_\bot(n\omega_l)L}(r_{mv})^{2}}  \Bigg|^2
  \left|  \frac{e^{iq_\bot(n\omega_l)L}-e^{inq_\bot(\omega_l)L}}{(nq_\bot(\omega_l)-q_\bot(n\omega_l))} + r_{mv}\frac{e^{iq_\bot(n\omega_l)L}-e^{i(nq_\bot(\omega_l)+2q_\bot(\omega_l))L}}{(nq_\bot(\omega_l)+q_\bot(n\omega_l))}\right|^2 \,. 
    \label{eq:int_normal_incidence_multiple}
\end{eqnarray}
Equation~\ref{eq:int_normal_incidence_multiple} formula is the equivalent of Eq.~\ref{eq:int_normal_incidence_simple} for the case of multiple reflections. We found numerically that in our experimental conditions, the multiple reflections do not play any important role.

 \end{widetext}
 
\subsection{Results for the propagation effects}

\begin{figure}
 \begin{centering}
 \includegraphics[width=0.8\columnwidth]{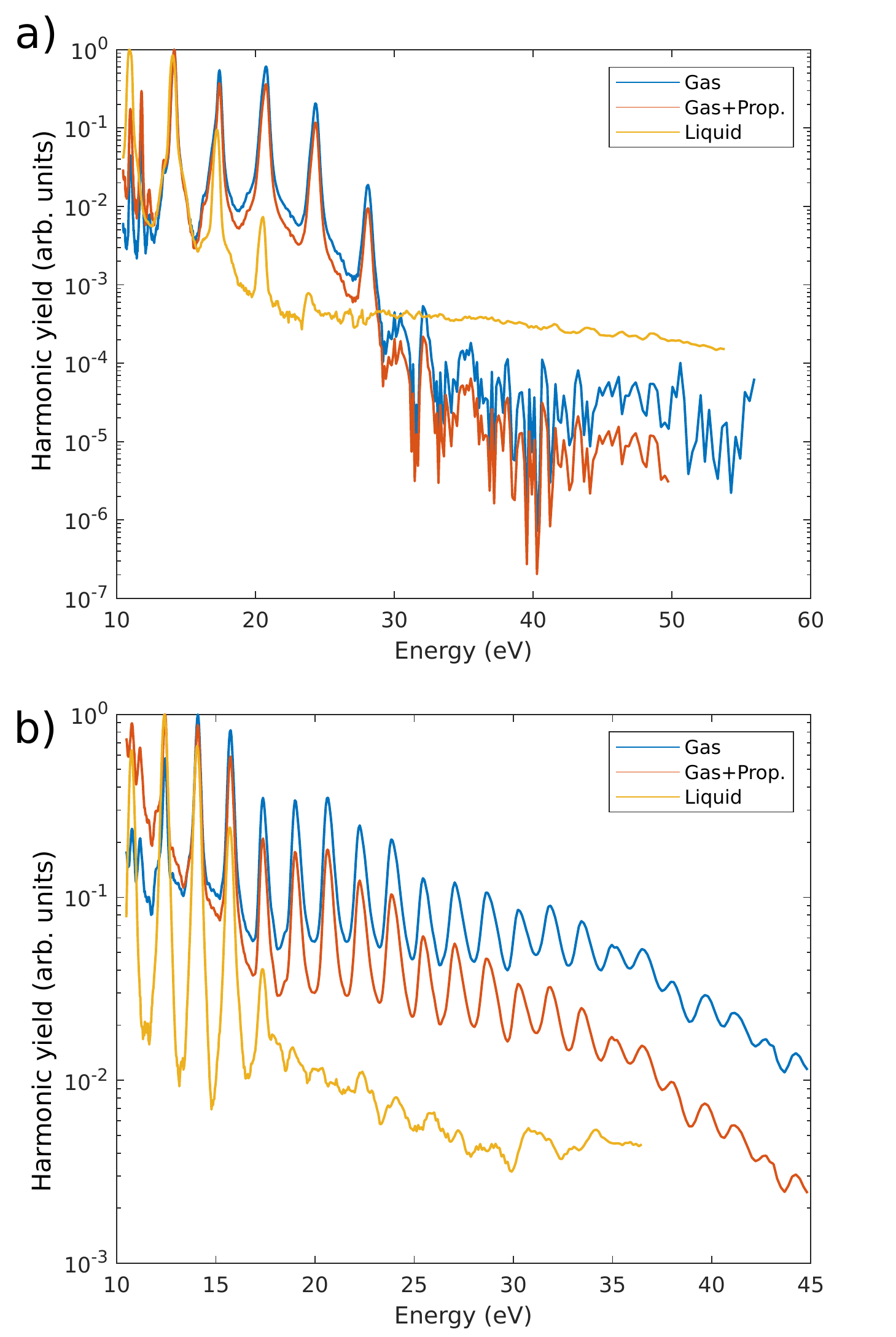}
 \caption{\label{fig:maxwell_prop} Effect of propagation for a 1$\mu m$ thick slab of liquid water. The emission is assumed to be given by the gas phase for the same laser intensity as for the liquid phase measurement. The effect of harmonic light propagations is computed here for (a) 800\,nm driving field, and (b) 1500\,nm driving field. In both cases, the change in the energy cutoff cannot be explained by the laser reabsorption.}
 \end{centering}
\end{figure}

We now employed the above model to simulate the effect of light propagation in water jet, and in particular to simulate the effect of re-absorption of the harmonic light by the jet itself. We considered a jet thickness of 1 $mu m$. We assumed that the laser generate an harmonic spectrum as in the gas phase at each possible position along the slab, and we employed the experimental dielectric function of water to model light-propagation effects~\cite{hayashi2015accurate}. Due to the reabsorption, we found that the total measured signal, that reaches the detector, builds up within the 200nm close to the exist surface. However, our results, shown in Fig.~\ref{fig:maxwell_prop} shows that the propagation effects, including phase-matching effects and light re-absorption, cannot explain the change in the energy cutoff. We also found (not shown) that the multiple reflections of either the incoming light or of the emitted harmonic light are not changing significantly the effect of the propagaition.

\section{Experimental results for density and propagation effects}
To isolate the effects of propagation and density liquid generated high harmonic spectra we experimentally measured jet thickness dependence of the harmonic cut-off and harmonic yield. If the cut-off energy is dependent on the propagation length through the liquid, a change in the jet thickness would show a change in the cut-off energy. The thickness for a flat-jet varies in the vertical direction. However, we have seen for Ethanol liquid (Figure 4(A) of manuscript) that the temperature also changes along the vertical direction. To isolate the effect of only the thickness variation we specifically choose H$_2$O and the vertical positions of(0.66-1.26 mm from top of jet). The thickness has been measured using a few-cycle 800 nm XUV-XUV high harmonic interference and verified through white light interferometry\cite{Yin_2020}.Figure S13(A) shows the fractional density change in percentage as a function of the jet position and the corresponding jet thickness. The density change has been calculated from the temperature variation as a function of the jet position, obtained from the calculations of \cite{chang2022} for H$_2$O flat-jet formed using 40 um nozzle diameter(used for the thickness measurements here).\cite{chang2022}. The corresponding water density has been interpolated from Table 17 of\cite{Patterson_1994}. It is observed that in this vertical scan range, while the thickness changes from 1000 nm to 500 nm the fractional change in density with respect to the 1000 nm data point is only about 0.04\%, thus for H$_2$O and the given jet positions we can clearly say, that whatever cut-off energy shift,in any, we observe should solely be because of the jet thickness variation.Figure S13 (B)  shows the harmonic spectrum of H2O measured for different jet thicknesses using 800 nm driving  wavelength. Each harmonic spectra has been normalized to the 5th harmonic (energy 7.76 eV) of 800 nm.  The Figure S13 (B) clearly indicates that the harmonic spectra are almost identical, irrespective of the jet thickness. To observe this carefully we plot Figure S13 (C) that shows the ratio of each harmonic signal (Normalized yield summed over an energy bin of 0.5 eV centered around each harmonic peak). The top x axis of Figure S13 (C) indicates the fraction change in density in percentage with respect to the data point at 1000 nm thickness (vertical position 0.66 mm). It is observed that the harmonic ratio  varies negligibly as a function of the jet thickness. This observation is in contrast to what we see for Ethanol: Figure 4(A) (Figure S14): where the fractional density is significantly large and we see a clear variation of the harmonic ratio as a function of the fractional density change. This further supports the microscopic origin of the cut-off energy shift, that is now clearly isolated to be a density effect as opposed to  an absorption effect of the liquid medium.

\begin{figure}
 \begin{centering}
 \includegraphics[width=\columnwidth]{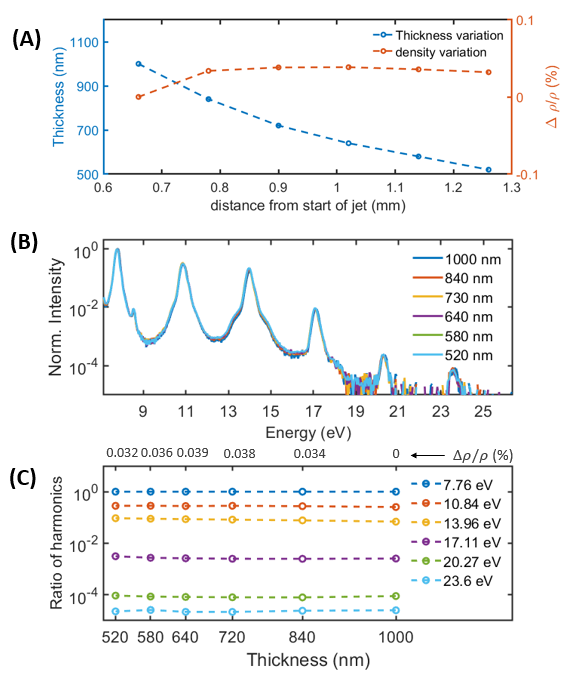}
 \caption{\label{fig:maxwell1} Effect of jet thickness on the cut-off energy of liquid harmonics of H$_2$O measured at 800\,nm wavelength. (A) Shows the fractional density change in percentage as a function of the jet position and the corresponding jet thickness obtained from \cite{Yin_2020}, \cite{chang2022} and \cite{Patterson_1994}.(B) The harmonic spectrum of H2O measured for different jet thicknesses using 800 nm driving  wavelength. Each harmonic spectra has been normalized to the 5th harmonic (energy 7.76 eV) of 800 nm.  (C) Ratio of each harmonic signal (Normalized yield summed over an energy bin of 0.5 eV centered around each harmonic peak). The top x axis of Figure S13 (C) indicates the fraction change in density in percentage with respect to the data point at 1000 nm thickness (vertical position 0.66 mm). }
 \end{centering}
\end{figure}

We compare these results to the harmonic yields observed for different spectrums presented in Fig. 4(A) of the manuscript, where we have normalized each spectrum to the 11 th harmonic of 1800 nm. Figure S14 (A) plots the normalized harmonic yields as a function of harmonic energy for the four different spectrums of Fig. 4(A)  where the first four energies correspond to the harmonics below cut-off (H11 H13 H15 H17) and the two highest energies(H19 and H21) correspond to harmonics above cut-off.  The harmonic yield has been computed by a summation of the harmonic signal within an energy bin of 0.5 eV centered on each harmonic peak. The measured signal is then normalized to the signal of harmonic 11.  Figure S14 (B) shows the harmonic ratios obtained for different density variations from the cluster calculations done for a laser wavelength of 900 nm. Comparing Figure S14 (B) with the Figure S14 (A) we see that the harmonics below the cut-off behave similar to what we expect from trajectory simulation, where reducing the fixed $\lambda_{MFP}$ values (equivalent to increasing density) systematically suppresses the harmonic yields, which essentially emphasizes the validity of the fixed MFP assumption. However for harmonics beyond cut-off (H19 and H21) the trend seems to change increasing in yield  for higher densities. This might involve contribution from the energy dependent MFP but harmonics beyond cut-off also have contributions from backscattering and nearest neighbor recombination as discussed previously. Thus isolating the contribution of just one effect is beyond the scope of this manuscript.

\begin{figure}
 \begin{centering}
 \includegraphics[width=0.8\columnwidth]{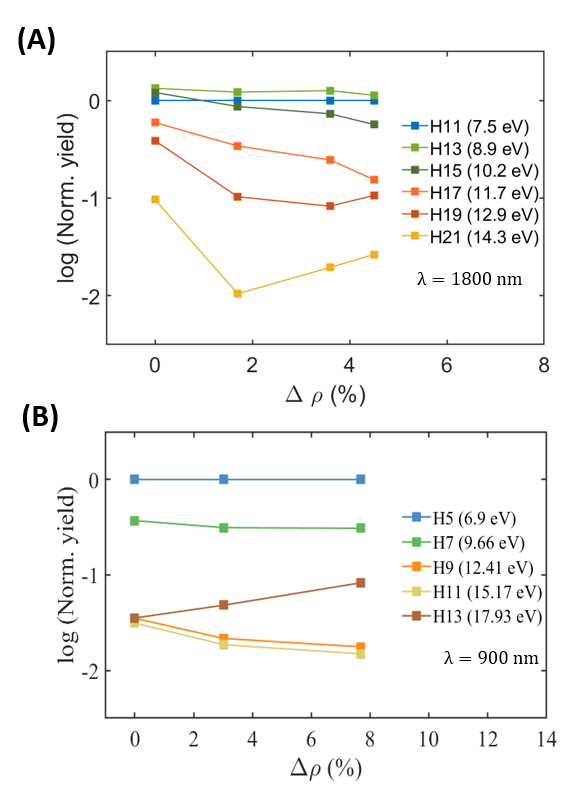}
 \caption{\label{fig:maxwell2} (A) Normalized harmonic yields as a function of harmonic energy for the four different spectrum of Fig. 4(A)  where the first three energy points correspond to the harmonics below cut-off (H11 H13 H15 and H17). (B) The harmonic ratios obtained for different density variations from the cluster calculations done for a laser wavelength of 900 nm and intensity of 4$\times$ 10$^{13}$ W/cm$^2$.}
 \end{centering}
\end{figure}

\section{Comparison of cut-off energy obtained from the semi-classical scattering model from nearest neighbor interaction and onsite interaction}

From the perspective of a semi-classical trajectory based model, the energy cutoff obtained from the recombination to the neighboring site is significantly  larger than for a wave packet recombining to its parent ion, because of a higher kinetic energy at the collision time. Thus, if recombination channels on neighboring molecules were playing a significant role the cutoff energy would have been almost double than that of the onsite recombination, which contradicts our observations. The figure below shows the comparison of the maximum kinetic energy as a function of laser wavelength (Figure S15 A) and as a function of the laser field (Figure S15 B) for the on-site recombination and nearest neighbor recombination as calculated from the semi-classical scattering model. In addition to the nearest neighbor interaction having higher cut-off than the onsite interaction, we also observe that for wavelengths below 900 nm the nearest neighbor interaction should ideally give a cut-off higher than the SFA (for the gas generated harmonics). But from the data presented in Figure S1 of the supplement, this is definitely not the case. Moreover we see that for the nearest neighbor interaction the maximum kinetic energy has a stronger dependence on the electric field than the onsite interaction, which is in contrast to our observations in Figure S1(D). 

\begin{figure}
 \begin{centering}
 \includegraphics[width=0.8\columnwidth]{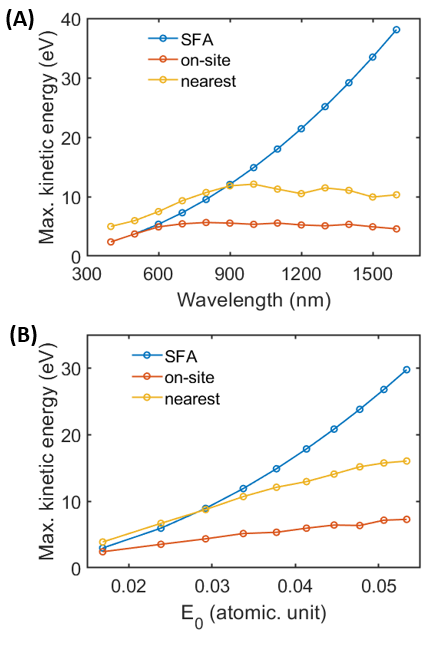}
 \caption{\label{fig:nearest_neighbor} Comparison of maximum kinetic energies for onsite recombination and nearest-neighbor recombination: (A) As a function of  wavelengths, calculated at an intensity of 5$\times$10$^{13}$ W/cm$^2$. (B) As a function of driving field for 1000 nm driving wavelength.}
 \end{centering}
 \end{figure}
 
\section{Change of the cutoff energy for elliptically polarized light}

In the case of elliptically-polarized light of a given polarization $\epsilon$, the semi-classical equation of motion of the electron, ionized at a time $t_i$, is given by
\begin{eqnarray}
 x(t) &=& \frac{qE}{m\omega^2} [\cos(\omega t_i)-\cos(\omega t) - \sin(\omega t_i) \omega(t-t_i)]\,,\nonumber\\
 y(t) &=& \frac{qE\epsilon}{m\omega^2} [\sin(\omega t_i)-\sin(\omega t)] + v_{0y}(t-t_i)\,,\nonumber\\
 v_x(t) &=& \frac{qE}{m\omega} [\sin(\omega t)-\sin(\omega t_i)] \,, \nonumber\\
 v_y(t) &=& -\frac{qE\epsilon}{m\omega} [\cos(\omega t)] +v_{0y} \,, \nonumber
\end{eqnarray}
where we imposed as initial condition that the trajectories start at $x(t_i)=y(t_i)=0$, with a non-vanishing velocity along the minor-axis direction ($y$), that guaranties that the electron returns to its parent ion ($y(t_r)=0$) at the return time. As in linearly-polarized case, the return time is given by the condition $x(t_r)=0$. This leads to $v_{0y}=\frac{qE\epsilon}{m\omega^2} \frac{\sin(\omega t_r)-\sin(\omega t_i)}{\omega (t_r-t_i)}$.\\
From these equation of motions, we considered only returning trajectories which travel a distance  $d=\int_{t_i}^{t_r} |\mathbf{v}(t)| dt$ smaller than an effective mean-free-path distance, and computed the corresponding gain in kinetic energy at the return time. This directly allows us to compute the change in cutoff energy for our model with respect to the driver ellipticity, as shown in the main text Figure ED1. We found that the energy cutoff decreases with increasing ellipticity. We note that if we instead would start from a zero velocity, we also obtain a decrease of he energy cutoff with the increasing ellipticity for small ellipticities, and an increase for large ellipticities, because of the unphysical starting point.
If we consider instead of the recombination to the parent ion the recollision with the nearest neighboring molecule, we obtain that the increasing ellipticity leads to an increase of the energy cutoff, which is the opposite trend as observed experimentally.

\bibliography{ref.bib,attobib.bib}